\documentclass[aps,10pt,showpacs,preprint,longbibliography, superscriptaddress, english, prl]{revtex4-2}
\usepackage[T1]{fontenc}
\usepackage[ruled,linesnumbered]{algorithm2e}
\usepackage{dsfont}
\usepackage{amssymb}
\usepackage{amsmath}

\usepackage{bm}

\usepackage{xspace}
\newcommand*{\eg}{\emph{e.g.}\@\xspace}
\newcommand*{\ie}{\emph{i.e.}\@\xspace}

\usepackage{enumitem}
\usepackage{graphicx}
\usepackage{subcaption}

\usepackage[colorlinks=true,urlcolor=blue,citecolor=blue,linkcolor=blue]{hyperref} 

\usepackage{amsmath}
\usepackage{mathtools}

\newcommand{\change}[1]{{{#1}}}
\newcommand{\changeT}[1]{{{#1}}}
\newcommand{\changeTT}[1]{{{#1}}}

\makeatletter

\begin{document}

\title{Deep generative modeling of the canonical ensemble with differentiable thermal properties}
\author{Shuo-Hui Li}
\email{shuohuili@ust.hk}
\thanks{These authors contributed equally to this work.}
\affiliation{Department of Physics, The Hong Kong University of Science and Technology, Hong Kong, China}

\author{Yao-Wen Zhang}
\thanks{These authors contributed equally to this work.}

\affiliation{Department of Physics, The Hong Kong University of Science and Technology, Hong Kong, China}

\author{Ding Pan}
\email{dingpan@ust.hk}
\affiliation{Department of Physics, The Hong Kong University of Science and Technology, Hong Kong, China}
\affiliation{ Department of Chemistry, The Hong Kong University of Science and Technology, Hong Kong, China}

\begin{abstract}
  It is a long-standing challenge to accurately and efficiently compute thermodynamic quantities of many-body systems at thermal equilibrium.
  The conventional methods, \eg, Markov chain Monte Carlo, require many steps to equilibrate.
  The recently developed deep learning methods can perform direct sampling,
  but only work at a single trained temperature point \change{and risk biased sampling}.
  Here, we propose a variational method for canonical ensembles with differentiable temperature, which gives thermodynamic quantities as continuous functions of temperature akin to an analytical solution.
  The proposed method \change{is a general framework} that works with \change{\emph{any}} tractable density generative model.
  At optimal, the model is \change{theoretically guaranteed to be the \changeT{\emph{unbiased}} Boltzmann distribution.} 
  We \change{validated} our method \change{by calculating} phase transitions in the Ising and XY models, \change{demonstrating} that our direct-sampling simulations are as accurate as Markov chain Monte Carlo, but more efficient.
  Moreover, 
  our differentiable free energy aligns closely with the exact one to the second-order derivative, indicating that
  the variational model captures the subtle thermal transitions at the phase transitions.
  This functional dependence on external parameters \change{is a fundamental advancement in combining the exceptional fitting ability of deep learning with rigorous physical analysis.}
\end{abstract}

\maketitle

The canonical ensemble is for a system in thermal equilibrium with a heat bath, where the temperature, as a principal thermodynamic variable, determines the probability distribution of states. 
Calculating canonical ensemble averages gives thermodynamic properties of condensed-matter many-body systems, \eg, interacting magnetic spins and solvated proteins. 
This calculation is closely related to the
partition function:
\begin{equation}
  \mathcal{Z} = \int \exp(-\beta E(\mathbf{x})) d\mathbf{x}, ~ F = -T\log \mathcal{Z},
  \label{eq:paritionFnExact}
\end{equation}
where $E$ is the energy function, $\beta$ is the inverse of the temperature $T$, and $F$ is the Helmholtz free energy.
For example, after obtaining $\mathcal{Z}$, the mean energy and the heat capacity can be readily calculated as derivatives of the free energy,
\begin{equation}
 \langle E \rangle = -\frac{\partial \log\mathcal{Z}}{\partial \beta}, ~ C_v = \beta^2\frac{\partial^2 \log\mathcal{Z}}{\partial \beta^2}.
  \label{eq:diffmeanEandCV}
\end{equation}

For most physical systems, it is impossible to analytically solve $\mathcal{Z}$:
an exact enumeration belongs to the class of \#P-hard \cite{partitionFn}.
In practical calculations, it is common to approximate $\mathcal{Z}$ numerically~\cite{trg, meanfield}.
However, even numerical approximation remains highly challenging, especially for systems undergoing phase transitions (PTs), where the partition function must account for a large number of metastable states.

Because of the complexity of $\mathcal{Z}$, sample-based methods are widely used in numerical calculations, among which the Markov chain Monte Carlo (MCMC) method holds prominence.
The MCMC method is an importance sampling technique giving the Boltzmann distribution of the canonical ensemble,
\begin{equation}
  p(\mathbf{x}) = \frac{\exp(-\beta E(\mathbf{x}))}{\mathcal{Z}}.
  \label{eq:boltzmannDis}
\end{equation}
It can provide the prevailing state configurations at a microscopic level, which are often unexplored in partition function approaches. This microscopic information is very valuable, \eg, in the identification of intermediate states in chemical reactions.
Macroscopic properties, such as thermodynamic quantities, can also be estimated by averaging the samples~\cite{SMtextbook}, namely, statistical averaging.
As MCMC has the unbiased estimation guarantee, this statistical averaging is the ensemble averaging.
An inherent limitation of MCMC is its efficiency. Because of the Markov chain structure of the rejection sampling, it takes multiple timesteps to equilibrate and to produce independent and identically distributed (i.i.d.) samples, which prohibits the MCMC from performing direct sampling~\cite{autocorrelation1}. This autocorrelation problem becomes even worse near PTs~\cite{autocorrelation1, autocorrelation2}.
Additionally, unlike the explicit temperature dependence in the partition function, the MCMC method typically requires more computational effort for results at multiple temperature points, further diminishing its overall efficiency.

\change{
Recently, a type of deep neural network model was introduced to address the efficiency limitations of MCMC: the tractable density generative model \cite{explicitDensity, prl, frankNoe}. 
As a generative model, it can directly produce one-shot i.i.d. samples from parameterized distributions, without requiring rejections.
Additionally, as an tractable density, it provides the exact probability of the sample~\cite{explicitDensity}.
}
Some of the models can also provide free energy estimation~\cite{prl, wuDianPRL}, which is deemed challenging for MCMC.
However, the limitation of temperature dependence persists, as separate models must be trained for each temperature point, forming an inefficient scheme similar to MCMC.
Beyond the efficiency issue,
this single-temperature training breaks the continuous temperature dependence of Boltzmann distribution.
This is reflected in biases, making the sampled distribution favor a few high-probability states and ignoring the vast meta-stable states, \changeTT{a phenomenon also known as the model collapse}.
In the case of MCMC, despite inefficiency, its unbiased estimation promises accurate results. However, deep learning models, being variational, may exhibit biases and lack physical fidelity.
When applying these models, one usually has to compensate with the aid of the MCMC rejection \cite{prl, PREfromRef}, annealing training \cite{wuDianPRL}, \change{or sophisticated neural network architectures~\cite{PRE2fromRef}.
As proposed below, biases can be mitigated by employing a temperature-dependent model that accounts for thermal effects in the distribution.
}

\paragraph{Variational temperature-differentiable method}
{Here, we introduce the variational temperature-differentiable (VaTD) method as a general framework for modelling the canonical ensemble in a continuous temperature range;
  by a minimal adaption of incorporating temperature as an input parameter, one can turn \emph{any} tractable density model into a VaTD model.
  After training with randomly sampled temperatures,
  this VaTD model can be an accurate direct-sampling model of the target Boltzmann distribution with temperature dependence.
  }Moreover, as the model is differentiable with temperature, the estimates, including the free energy and the partition function, can be differentiated similarly to an analytical solution.

First, considering a single temperature, one can minimize the reverse Kullback-Leibler divergence (KLD)~\cite{KLD} to optimize an tractable density generative model, \ie,
\begin{equation}
    D_{KL}(q_{\theta}||p) = \mathop{\mathbb{E}}_{\mathbf{x}\sim q_{\theta}} \log\frac{q_{\theta}(\mathbf{x})}{p(\mathbf{x})}
                          = \log\mathcal{Z} + \mathop{\mathbb{E}}_{\mathbf{x}\sim q_{\theta}} \left[\log q_{\theta}(\mathbf{x}) + \beta E(\mathbf{x})\right],
  \label{eq:KLD}
\end{equation}
where $q_{\theta}$ is the model distribution,
and $p$ is the Boltzmann distribution at \changeT{this fixed} temperature (Eq.~(\ref{eq:boltzmannDis})).
After optimization, the model $q_{\theta}$ is an approximation of $p$.
Additional, because the $D_{KL} \ge 0$, the statistical average, $\mathbb{E}_{\mathbf{x}\sim q_{\theta}} [\log q_{\theta}(\mathbf{x}) + \beta E(\mathbf{x})]$, has a variational lower bound at $-\log\mathcal{Z}$,
it can thus be used as an estimate of the partition function in Eq.~(\ref{eq:paritionFnExact})~\cite{prl, wuDianPRL}.
This has been successfully demonstrated in various applications \cite{equivariantFlowLattice, lattice2, freePart1, freePart2, lw2023PRL}.
\change{
However, since the actual Boltzmann distribution is temperature-dependent, training schemes that optimize for a single temperature may become biased toward the prevailing states at that temperature. Additionally, as shown in Eq.~(\ref{eq:KLD}), the model is optimized based on its own samples, which can lead to persistent bias.
}

\change{
  In the VaTD training, we propose to approximate the Boltzmann distribution by training in a continuous temperature range.
  \changeT{Varying temperatures can reveal hidden local minima, a concept akin to parallel tempering in MCMC, where high temperatures are applied to surmount energy barriers.}
  This motivates us to derive a temperature-dependent loss function from the variational definition of the Boltzmann distribution,
  \ie,
   the distribution minimizes free energy across a continuous temperature range:
  }
\begin{equation}
\begin{split}
  \min_{q_{\theta}} \sum_{\{\mathbf{x}\}} \left[1/\beta \cdot q_{\theta}(\mathbf{x}, \beta)\log q_{\theta}(\mathbf{x}, \beta) + q_{\theta}(\mathbf{x}, \beta)E(\mathbf{x})\right],
  \\
  \text{s.t.} ~ \sum_{\{\mathbf{x}\}} q_{\theta}(\mathbf{x}, \beta) = 1, ~ \forall \beta.
  \label{eq:optimBoltzmann}
\end{split}
\end{equation}
Using the Lagrange multiplier, one can see that the Boltzmann distribution in Eq.~(\ref{eq:boltzmannDis}) is the solution \cite{SM}.

\change{
  Having temperature as an input parameter, the VaTD models can be used as the variational $q_{\theta}(\bm{x}, \beta)$.
  \changeT{This allows Eq.~(\ref{eq:optimBoltzmann}) to be formulated in a form that resembles Eq.~(\ref{eq:KLD}), enabling straightforward deep-learning training.}
  }
As tractable density models, VaTD models are naturally normalized.
We then turn the summation into estimation, and the minimum at each $\beta$ into the minimum of an estimation over uniformly distributed $\beta$.
Thus, the following loss function is equivalent to the free energy minimization in Eq.~(\ref{eq:optimBoltzmann}):
\begin{equation}
  \mathcal{L} = \mathop{\mathbb{E}}_{\substack{\beta \sim \text{U} \\ \mathbf{x}\sim q_{\theta}(\bm{x} , \beta)}} \left[\log q_{\theta}(\mathbf{x}, \beta) + \beta E(\mathbf{x})\right] = \mathop{\mathbb{E}}_{\beta \sim \text{U}} \left[-\log\bar{\mathcal{Z}}(\beta)\right].
  \label{eq:estPartFn}
\end{equation}
The stochastic gradient descent (SGD) \cite{SGD1, SGD2} can be performed to optimize it.
Gradients can be obtained from the computational graph and automatic differentiation, which are widely used in deep learning \cite{autodiff}.
\change{As ensured in Eq.~(\ref{eq:optimBoltzmann}), optimizing this loss drives the model toward the unbiased Boltzmann distribution. Hence, after training, the model can accurately learn the actual Boltzmann distribution along with its temperature dependence}.

\changeT{Besides its thermodynamic origin,
this loss function also represents an integration of the reverse KLD from Eq.~(\ref{eq:KLD}) over a continuous temperature range.}
Consequently, this training doesn't require any dataset, as the sample is drawn from a generative model by direct sampling.
Another benefit is that one gets a temperature-differentiable estimate of the free energy, $\log\bar{\mathcal{Z}}(\beta)$.
So, its derivatives with respect to temperature,  \eg, the mean energy and heat capacity in Eq.~(\ref{eq:diffmeanEandCV}), can be readily calculated using automatic differentiation.
\change{
  As shown in the SI~\cite{SM},
  this is more time efficient than the sample-based approach.
  }
Other thermodynamic quantities, \eg,  magnetization, can be estimated by statistical averaging of directly sampled batches, and they are also differentiable functions of temperature.

One concern is that $q_{\theta}$ may not be flexible enough to cover the whole space of probability distribution, which may result in sub-optimal solutions for Eq.~(\ref{eq:optimBoltzmann}) and miss the Boltzmann distribution.
However, 
as guaranteed by the universal approximation theorem \cite{universalnet},
a sufficiently large neural network model can represent any distribution.
\change{
Since VaTD training is applicable to any tractable density model, it is easy to apply more advanced model structures. 
The VaTD training does not require a redesign of the model. Instead, it simply integrates temperature as an input, thus preserving the model’s fitting capabilities.
In contrast to previous works that involve constructing specialized models designed to consistently follow the Boltzmann distribution (e.g., \cite{NoeTemperatureFlow}), our method serves as a general
training framework for \emph{any} tractable density model. Additionally, in the VaTD training, the model begins as a random distribution and is subsequently optimized to approximate the unbiased Boltzmann distribution
}

In performing SGD, it is worth noting that the gradients of Eq.~(\ref{eq:estPartFn}) are not trivial.
As the model parameters $\theta$ are used in the expectation operator, the nice form of statistical averaging will be lost in the gradients.
To overcome this, two numerical methods, namely reparameterization and REINFORCE estimation, have been proposed \cite{reparamAndreinforce}.
They enable the swapping of the differentiation operation with the expectation operation, thereby converting the derivative calculations of statistical averages into the statistical averaging of derivatives.
The same problem reoccurs when calculating the derivative of the estimated thermal differentiable function.
This can be solved again by using these two methods.
Typically, only the first-order forms of these two methods are presented for training.
We extend them to the second order in SI~\cite{SM}.
It is worth noting that higher-order generalizations are feasible as long as the neural network remains continuous.

\begin{figure*}[tb]
\centering
\includegraphics[width=0.5\textwidth]{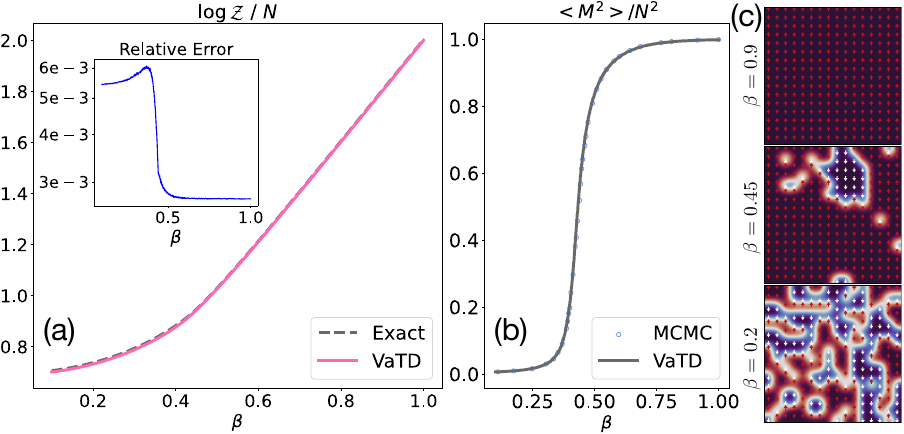}
\caption{
  (a) The exact and estimated free energy of the 2D Ising model on a $16\times 16$ square lattice with PBC, and the relative error between the two (inset). The temperature factor ($-T$) is removed from the free energy($-T \log \mathcal{Z}$) for better comparison.
  (b) The estimated squared magnetization of the Ising model.
  (c) State configurations directly sampled from the trained model at three temperatures, with background color as a continuous interpolation of the discrete site for better visualization.
}
\label{fig:ising}
\end{figure*}

\begin{figure*}[tb]
\centering
\includegraphics[width=0.5\textwidth]{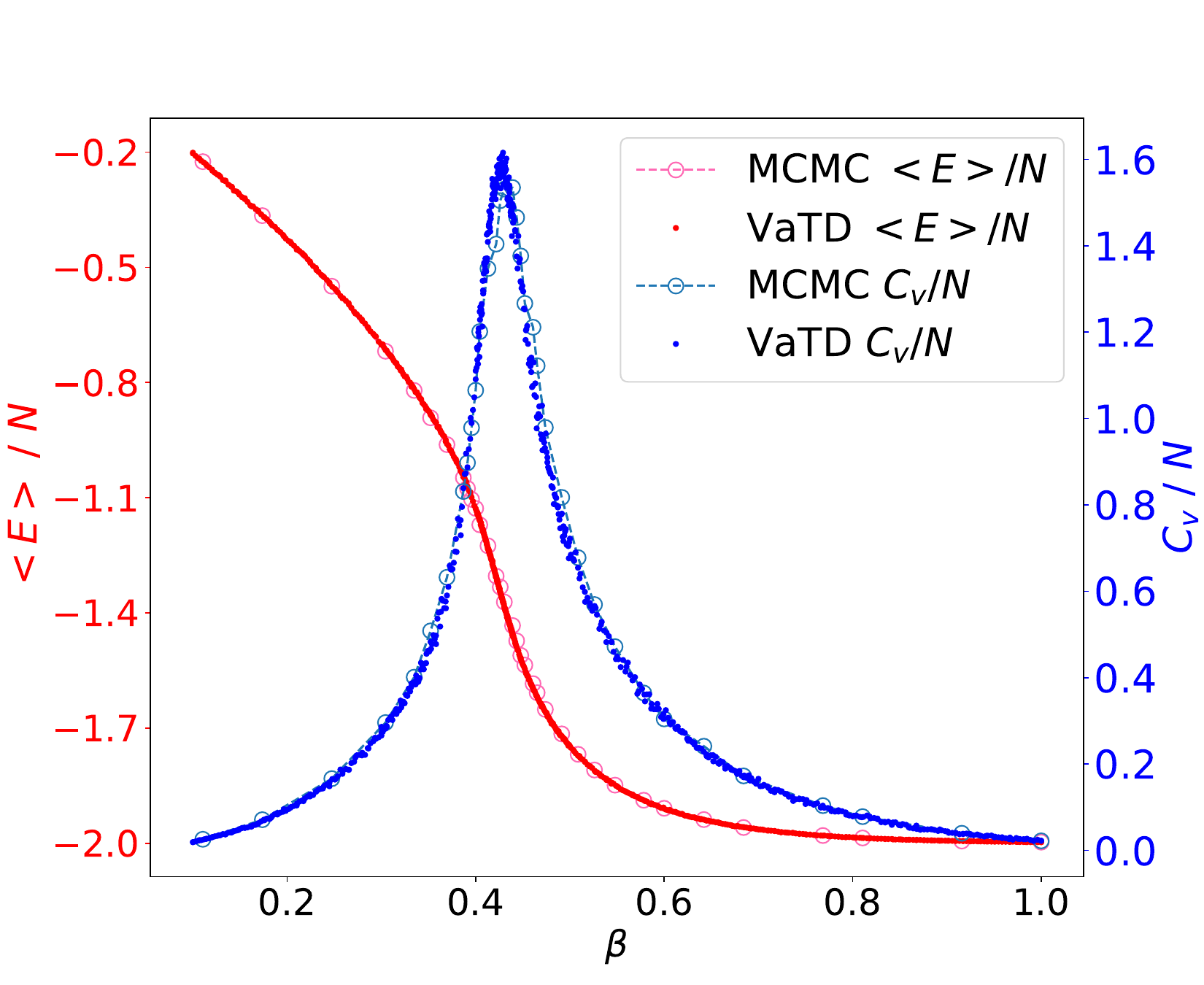}
\caption{
  The mean energy and heat capacity of the Ising model, estimated using the differentiation of the free energy, compared with the results obtained from the MCMC simulation.
}
\label{fig:ising2}
\end{figure*}

\paragraph{Numerical experiments}
In the numerical experiments, we 
validated our method using two famous statistical models: two-dimensional Ising and XY models. 
The former one uses a discrete-variable model, PixelCNN \cite{pixelRCNN},
as the tractable density model used in the VaTD method, while the latter one uses a continuous-variable model, the normalizing flow (NF).

Our 2D Ising model is on a $16\times 16$ square lattice with periodic boundary conditions (PBCs).
The energy function is given by $E(\mathbf{x}) = -\sum_{<i,j>} x_ix_j$, where $<i, j>$ means the two spins are nearest neighbors on the lattice.
Variables of Ising are discrete, with each variable $x_i$ taking a value of either $1$ or $-1$, representing spin-up or spin-down, respectively.

For Ising, we incorporated the temperature \change{as an input} into the PixelCNN~\cite{pixelRCNN} as the VaTD model. PixelCNN is an autoregressive generative model for discrete-variable distributions.
Our VaTD PixelCNN's probability distribution is decomposed into the product of a series of conditional probability distributions, \ie, $q_{\theta}(\mathbf{x}, \beta) = \prod_{i}q_{\theta}(x_i | x_0,\cdots,x_{i-1};\beta)$.
The conditional distribution we chose is the Bernoulli distribution, $x_i \sim B(1, p)$, where $p$ is the probability of $x_i$ being in the spin-up state, as obtained from a residual network \cite{resnet} given the preceding variables $\{x_0,\cdots,x_{i-1}\}$ and the temperature parameter $\beta$.
To have a unified input form for the residual network, we repeated $\beta$ to match the size of $\{x_0,\cdots,x_{i-1}\}$, and then concatenated them along the channel dimension.
The PixelCNN model generates variables sequentially, one at a time. During each step, a variable is stochastically sampled from the Bernoulli distribution. The conditional causality is established through the dependence of the Bernoulli parameter $p$ on the preceding variables.
The model was trained in a temperature range of $\beta \in [0.05, 1.2]$. 
To eliminate the $\mathbb{Z}_2$ symmetry, we fixed the first spin to be spin-up.

After training, the VaTD PixelCNN model can perform direct sampling, and the sampled batches were used for statistical averaging.
In Fig.~\ref{fig:ising}(a), we estimated the free energy as a continuous function of temperature using Eq.~(\ref{eq:estPartFn}).
As the 2D Ising can be analytically solved~\cite{isingExact}, we compared our numerical results with the exact free energy at each temperature point in Fig.~\ref{fig:ising}(a), and found that the relative errors are small, with the maximum happening near the PT region where the system exhibits long-range correlations.
Similarly, we also calculated the squared magnetization, which is in excellent agreement with the result obtained by the unbiased MCMC simulation, as shown in Fig.~\ref{fig:ising}(b).
This indicates that our statistical averages from direct sampling follow the actual ensemble averages closely, similar to MCMC but more efficient.
The sampled configurations also provide valuable insights into the microscopic behaviors of the system, and demonstrate that 
the underlying physical transition is successfully captured.
In Fig.~\ref{fig:ising}(c), we observe that, at lower $\beta$, the uniform spin direction gradually collapses. 
Hence, our trained neural network model undergoes the same microscopic thermal transition as the target Ising model.

To further demonstrate that our model \change{unbiasedly} captures the subtle thermal transition effects near the PTs, we computed the mean energy and heat capacity, which are derivatives of the free energy as shown in Eq.~(\ref{eq:diffmeanEandCV}), using automatic differentiation on the estimated free energy. We then compared with the unbiased MCMC results in Fig.~\ref{fig:ising2}.
Overall, our results agree very well with those from the MCMC simulation, with only slight deviations observed near the PT region.
This implies that the estimated free energy aligns closely with the exact one to the second-order derivative.
Moreover, compared with statistical averaging using direct sampling, this free energy derivative estimation of mean energy and heat capacity has advantages in efficiency and accuracy
(see SI~\cite{SM} for detailed comparisons).

\begin{figure*}[tb]
\centering
\includegraphics[width=0.5\textwidth]{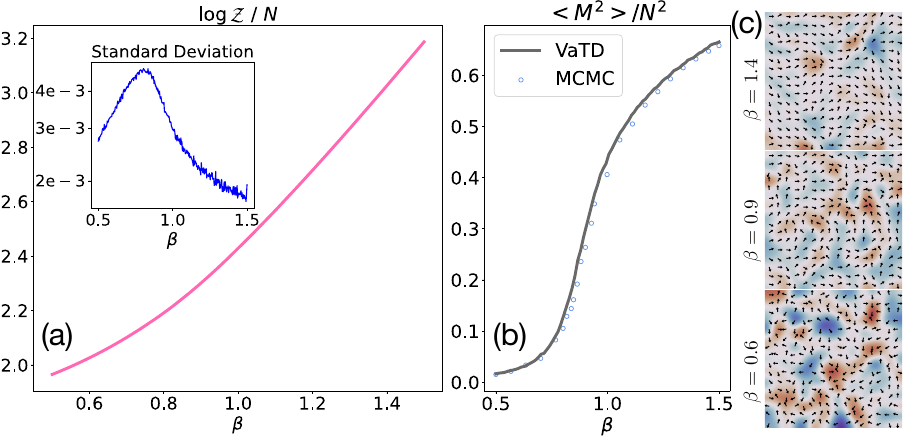}
\caption{
  (a) The estimated free energy and its standard deviation (inset) of the 2D XY model on a $16\times 16$  square lattice with PBC. The temperature factor ($-T$) is removed from the free energy($-T \log \mathcal{Z}$) for better comparison.
  (b) The estimated squared magnetization of the XY model. 
  (c) State configurations directly sampled from the trained model at three temperatures, the background color represents the vorticity.
}
\label{fig:xy}
\end{figure*}

\begin{figure*}[tb]
\centering
\includegraphics[width=0.5\textwidth]{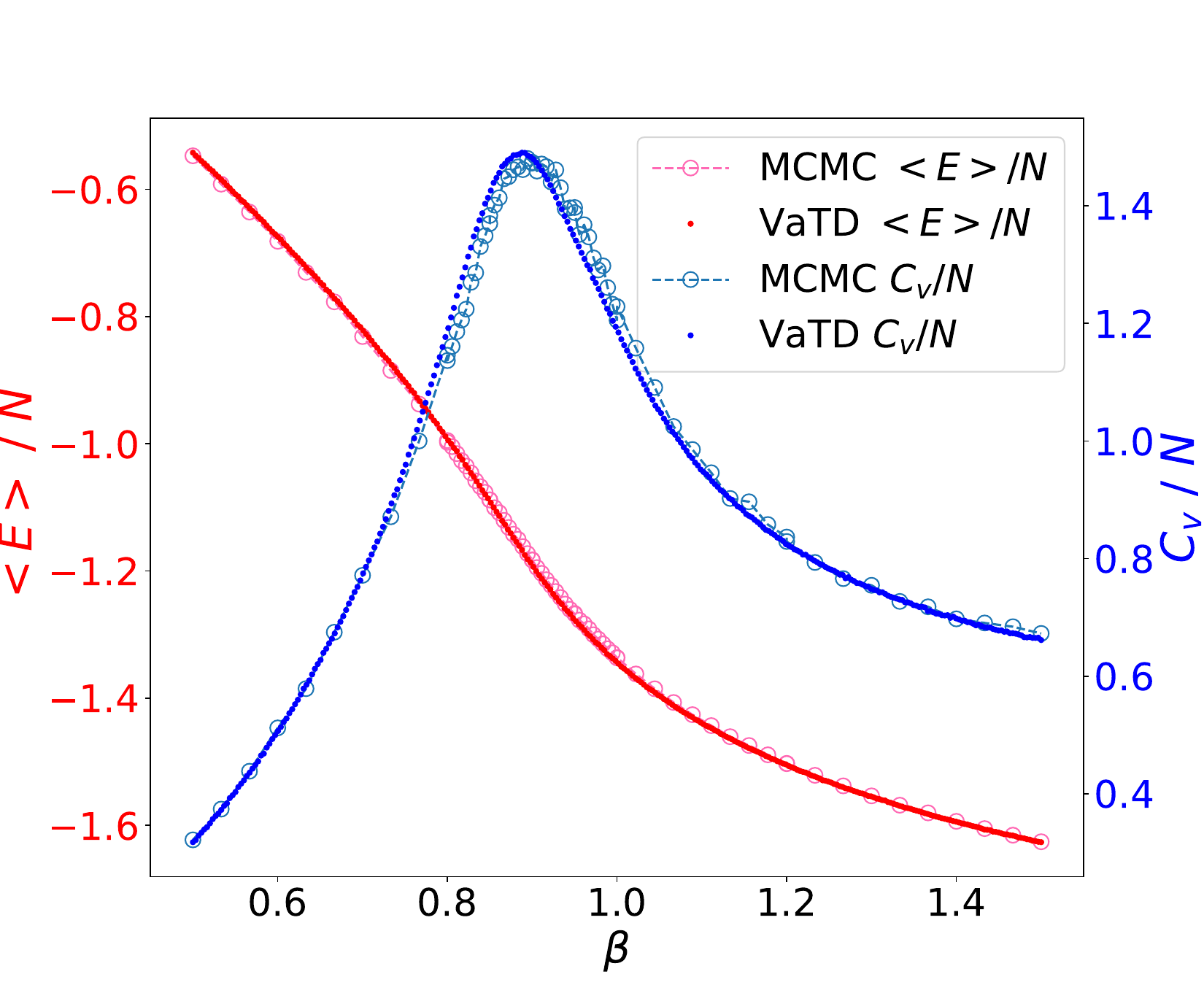}
\caption{
  The mean energy and heat capacity of the XY model, estimated using the differentiation of the free energy, compared with the results obtained from the MCMC simulation.
}
\label{fig:xy2}
\end{figure*}

Next, the XY model is more complex than the Ising model.
As an example of the Berezinskii-Kosterlitz-
Thouless transition~\cite{xyAndKT1,xyAndKT2}, the XY model finds practical applications in explaining various systems, including thin $^{4}\mathrm{He}$ films~\cite{XYhe}, and quasi-2D layered superconductors~\cite{XYsc}.
The energy function of the XY model is $E(\mathbf{x}) = - \sum_{<i,j>} \cos(x_i-x_j)$, where $x_i$ and $x_j$ are angles in the continuous range of $[-\pi, \pi]$.
Similar to the Ising case, we also used a 2D $16\times 16$ square lattice with PBC.

For XY's continuous variables, we incorporated the temperature \change{as an input} into a NF model~\cite{normalizingflow1, normalizingflow2, pwlinearAndquadratic, nsf, cubic} as the VaTD model.
In this VaTD NF model, the samples were initially drawn from a prior distribution, which is easy to sample from.
We chose the von Mises distribution, $\text{VM}(\mu, \kappa)$ where $\mu$ controls the mean and $\kappa$ controls the variance~\cite{vonMisesSample2, vonMisesSample1}, as the prior distribution.
To have a temperature-dependent prior distribution, we modified $\kappa$ as $\kappa_\beta = \kappa \cdot \beta$.
As VMs were sampled by an acceptance-rejection method, a special reparameterization should be used to optimize $\mu$ and $\kappa$ \cite{acptReparam, SM} (see the SI).
The sample $\bm{z}$ from the VM prior distribution was then transformed using parameterized invertible transformations, \ie, $\bm{x} = f_{\theta}(\bm{z}, \beta)$.
To incorporate temperature, we again repeated $\beta$ and padded it with the sample variables $\bm{z}$ in the channel dimension, which is similar to the Ising case.
The effects of these invertible transformations, besides changing the sample variables, also introduce probability changes to the VM prior distribution.
These distribution changes are measured by the determinant of the Jacobian matrix \cite{normalizingflow1}.
The invertible transformation we chose is the piece-wise cubic spline transformation \cite{cubic}.
This kind of transformation is designed for continuous variables in a closed range.
Then, the final probability of a sample is the product of the prior probability and the determinant of the Jacobian, \ie,
$q_{\theta}(\mathbf{x}, \beta) = \text{VM}(\bm{z}|\mu, \kappa_{\beta}) \cdot |\text{det} \frac{\partial \bm{z}}{\partial \mathbf{x}}|$ where $\bm{x} = f_{\theta}(\bm{z},\beta)$.
This model was trained in the temperature range of $\beta\in [0.4, 2.0]$. 
The $U(1)$ symmetry was eliminated by pinning the first spin to $0$.

After training, we performed direct sampling and used the obtained batches to perform statistical averaging.
Fig.~\ref{fig:xy}(a) shows the estimated free energy as a continuous function of temperature.
As XY is not exactly solvable, we calculated the standard deviation (STD) to evaluate the estimation quality.
According to Eq.~(\ref{eq:boltzmannDis}), a perfectly trained model would produce uniform free energy estimates regardless of $\bm{x}$, so the small STDs in Fig.~\ref{fig:xy}(a) indicate accurate \change{alignment with the target Boltzmann distribution}.
Then, to demonstrate our statistical averages accurately follow the ensemble averages, we compared our direct-sampling simulation results with the unbiased MCMC results;
Fig.~\ref{fig:xy}(b) shows the estimated squared magnetization, 
in excellent agreement with the MCMC results.
To study the microscopic behavior, we plotted the sampled configurations at different temperatures in Fig.~\ref{fig:xy}(c), which clearly shows the microscopic changes occurring near the PT. Specifically, with decreasing temperature, the vortex density decreases, and a quasi-long-range order emerges.
This demonstrates an accurate capture of the underlying physical transition.

Furthermore, in Fig.~\ref{fig:xy2}, we compared the mean energy and heat capacity obtained by the MCMC simulation and the differentiation of the estimated free energy. 
In comparison to the MCMC result, we observed only minor deviations around the PT, indicating a good fit of the free energy to the second-order derivative and a good capture of the subtle PT transition.
Numerical results in SI also show that this free energy derivative estimation of mean energy and heat capacity gives better convergence than the statistical averaging with direct sampling~\cite{SM}.
\change{Moreover, the derivative approach expresses heat capacity as a mean value over a batch of independent estimates, as shown in Eq.~(\ref{eq:diffmeanEandCV}).
This results in faster convergence than the energy deviations used in sample-based approaches; we have a detailed theoretical and numerical analysis in the SI~\cite{SM}.}
In contrast to the one-by-one generation process of PixelCNN, the NF model is significantly faster due to its all-to-all style of generation. 
\changeT{Including training time, the NF model matches Hamiltonian Monte Carlo in estimating $18$ temperatures at  a convergence accuracy of $10^{-3}$, and it becomes faster as the number of temperatures increases.}
Additionally, the low STD allows us to estimate the heat capacity using the first-order derivative, which offers a significant speedup and reduces the memory requirement. We give a detailed description of this in SI~\cite{SM}.
\changeT{
The proposed method is also effective for the XY model with lattice sizes up to 258 × 258. Larger lattices can be accommodated with increased GPU memory or additional GPU cards.
}

\paragraph{Conclusion and Outlook}
Here, we present a novel method for 
variational modeling of the canonical ensemble in a continuous temperature range.
\change{
  In the proposed method,
  any tractable density generative model can be turned into a VaTD model by directly including temperature as a part of the input.
  By converting the variational definition of Boltzmann distribution into a temperature-dependent loss, we propose the VaTD training framework to unbiasedly optimize the model.
  }
This allows thermodynamic quantities, such as free energy and partition function, to be differentiable functions of temperature.
Resembling analytic solutions, this differentiable free energy gives a concise way of expressing thermodynamic quantities as its derivatives.
Our method is as accurate as the unbiased MCMC method but more efficient.
Moreover, as a single model with temperature dependence, its direct-sampling estimation remains valid across a continuous temperature range. In contrast to previous deep generative models that optimize at a single temperature, our method preserves the essential temperature dependence. As a result, there is no need for training additional models at discrete temperature points, and a single model can reliably capture subtle thermal effects, even at PTs.

\change{As a general training framework, with minor modification, we expect to have immediate practical applications 
by adapting our method to these popular variational deep generative models at a single temperature~\cite{equivariantFlowLattice, lattice2, freePart1, freePart2, lwQuantumDots, lwelegas, lw2023PRL, prr}.}
Moreover, our proposed method can be seen as a generalization of the variational mean-field method, which consists of multiple conditional parameters. While our current implementation focuses on temperature, it opens up intriguing possibilities for exploring other parameters, such as pressure and external magnetic field.
In numerical experiments, we mainly applied our method to lattice models, but it is also viable to 
extend its application to more realistic systems with many meta-stable states, \eg, atomistic simulations.

\paragraph{Acknowledgments}
\begin{acknowledgments}
We thank Lei Wang, Wen-Qing Xie, Chu Li, Tao Li, and Jun-Ting Yu for many useful discussions.
S.-H.L. and D.P. acknowledges support from Hong Kong Research Grants Council (RGC) (GRF-16302423, GRF-16301723).
D.P. acknowledges support from the Croucher Foundation through the Croucher Innovation Award, National Natural Science Foundation of China (NSFC) through the Excellent Young Scientists Fund (22022310), and the NSFC/RGC Joint Research Scheme (N\_HKUST664/24).
Part of this work was carried out using computational resources from the National Supercomputer Center in Guangzhou, China, and the X-GPU cluster supported by the RGC Collaborative Research Fund C6021-19EF.
\end{acknowledgments}

\emph{Data availability}—The code implementation is publicly
available at \cite{Code}.

\bibliography{ref}
\appendix

\clearpage
\pagebreak
\widetext
\begin{center}
  \textbf{Supplementary Information: Deep generative modelling of canonical ensemble with differentiable thermal properties}
\end{center}

\setcounter{equation}{0}
\setcounter{figure}{0}
\setcounter{table}{0}
\setcounter{algocf}{0}

\renewcommand{\theequation}{S\arabic{equation}}
\renewcommand{\thefigure}{S\arabic{figure}}
\renewcommand{\thetable}{S\arabic{table}}
\renewcommand{\thealgocf}{S\arabic{algocf}}

\section{Introduction to  tractable density Generative Model}

\change{
The tractable density generative model is a special kind of generative model.
Generative model refers to deep neural network models that can generate samples in a high-dimensional space.
This is different from the discriminative model, where usually only a one-dimensional output is given as the category or yes-or-no label of the input data.
Examples of the generative model are variational autoencoders (VAE) \cite{vae} and generative adversarial networks (GAN) \cite{gan}.
Most of generative models are viewed as probability distributions, and the generation of samples is a one-shot i.i.d. sampling process.

The tractable density generative model is different from general generative models.
As an tractable density, it can give the exact probability of the sample.
This is usually achieved with specially designed structures or transformations of the neural network model.
As two examples, the two most used tractable density generative models are the normalizing flow model \cite{normalizingflow1, normalizingflow2, pwlinearAndquadratic, nsf, cubic} and variational autoregressive models (\eg, PixelCNN~\cite{pixelRCNN}).
We have introduced the two in the numerical experiments.
With this tractable density feature, one can use the model as a sampler inside MCMC algorithm~\cite{prl, PREfromRef}.
This is not possible with other types of generative models, as the Metropolis-Hasting rule requires knowledge of the proposal probability. 
Furthermore, tractable density models trained with reverse KLD (Eq.~\ref{eq:KLD}) will allow the estimation of partition function and free energy, as proposed in \cite{prl, wuDianPRL}.
}

\section{Boltzmann Distribution as the Variational Minimization of the  Free Energy}

Given a probability distribution $q(\mathbf{x})$ and an energy function $E(\mathbf{x})$, the corresponding free energy is
\begin{equation}
  F = \langle E \rangle - TS = \sum_{\{\mathbf{x}\}} \frac{1}{\beta} q(\mathbf{x}) \log q(\mathbf{x}) + q(\mathbf{x}) E(\mathbf{x}).
\end{equation}
As a probability distribution, $q(\mathbf{x})$ has the following constraint,
\begin{equation}
  \sum_{\{\mathbf{x}\}} q(\mathbf{x}) = 1.
\end{equation}
Considering the probability distribution $q^{*}$ that minimizes the free energy at all temperatures, then this $q^{*}$ should be the solution to the following optimization problem,
\begin{equation}
  \begin{aligned}
    \label{eq:apFreeE}
    \min_{q(\mathbf{x})} ~ &\frac{1}{\beta} \Big[\sum_{\{\mathbf{x}\}}q(\mathbf{x}) \log q(\mathbf{x}) + \beta q(\mathbf{x}) E(\mathbf{x})\Big],\\
    \text{s.t.}& ~ \sum_{\{\mathbf{x}\}} q(\mathbf{x}) = 1,~ \forall \beta.
  \end{aligned}
\end{equation}
This optimization can be analytically solved via the Lagrange multiplier method \cite{optimbook}, which gives the Boltzmann distribution as a function of temperature, \ie,
\begin{equation}
  {q}^{*}(\mathbf{x}) = \frac{\exp (-\beta E(\mathbf{x}))}{\sum_{\{\mathbf{x}\}}\exp(-\beta E(\mathbf{x}))} = \frac{\exp (-\beta E(\mathbf{x}))}{\mathcal{Z}}.
\end{equation}

To solve this in a variational way, we use a variational model $q_{\theta}$.
Then, because the requirement of the minimum at every $\beta$ point implies the minimum of an integration over all $\beta$ points, we can turn Eq.~(\ref{eq:apFreeE}) into
\begin{equation}
  \begin{aligned}
    \label{eq:apFreeEint}
    \min_{\theta} \int &\frac{1}{\beta} \Big[\sum_{\{\mathbf{x}\}}q_{\theta}(\mathbf{x}) \log q_{\theta}(\mathbf{x}) + \beta q_{\theta}(\mathbf{x}) E(\mathbf{x})\Big] d\beta,\\
    \text{s.t.}& ~ \sum_{\{\mathbf{x}\}} q(\mathbf{x}) = 1.
  \end{aligned}
\end{equation}
This can also be interpreted as an integration over an equal-probability distribution of $\beta$, which can be estimated via a uniform sampling, \ie,
\begin{equation}
  \begin{aligned}
    \label{eq:apFreeEest}
    \min_{\theta} \mathop{\mathbb{E}}_{\beta\sim U} &\frac{1}{\beta} \Big[\sum_{\{\mathbf{x}\}}q_{\theta}(\mathbf{x}) \log q_{\theta}(\mathbf{x}) + \beta q_{\theta}(\mathbf{x}) E(\mathbf{x})\Big], \\
    \text{s.t.}& ~ \sum_{\{\mathbf{x}\}} q(\mathbf{x}) = 1.
  \end{aligned}
\end{equation}
Using an tractable density model, one can directly sample $\bm{x}$ to perform statistical averaging as an estimation of the summation over $\{\bm{x}\}$, 
and the probability normalization condition is automatically satisfied, so we get
\begin{equation}
    \label{eq:apFreeEest2}
    \min_{\theta} \mathop{\mathbb{E}}_{\beta\sim U} \frac{1}{\beta} \Big[\mathop{\mathbb{E}}_{\bm{x}\sim q_{\theta}(\cdot,\beta)}\log q_{\theta}(\mathbf{x}) + \beta E(\mathbf{x})\Big],
\end{equation}
which is the Eq.~(\ref{eq:estPartFn}) in the main text.

\section{Reparameterization and REINFORCE Estimation}

In this section, we introduce the two methods for calculating the derivative of expectation operation. In the main text, these methods are used to compute the optimization gradients and the derivatives of free energy.

\subsection{First- and second-order reparameterization}

The reparameterization introduces a new random variable $\epsilon$ and rewrites the original expectation into an expectation over the unparameterized distribution of $\epsilon$, \ie,
\begin{equation}
  \mathop{\mathbb{E}}_{\mathbf{x}\sim q_{\theta}(\mathbf{x})}f(\mathbf{x}; \theta) = \mathop{\mathbb{E}}_{\epsilon \sim \pi(\epsilon)}f(h_{\theta}(\epsilon);\theta),~ \mathbf{x} =h_{\theta}(\epsilon).
\end{equation}
The first- and second-order derivatives can be reformulated as
\begin{equation}
  \begin{split}
    \nabla_{\theta}\mathop{\mathbb{E}}_{\mathbf{x}\sim q_{\theta}(\mathbf{x})}f(\mathbf{x}; \theta) = & \nabla_{\theta}\mathop{\mathbb{E}}_{\epsilon \sim \pi(\epsilon)}f(h_{\theta}(\epsilon); \theta)\\
      = &\mathop{\mathbb{E}}_{\epsilon \sim \pi(\epsilon)} \nabla_{\theta}f(h_{\theta}(\epsilon); \theta), \\
    \nabla^2_{\theta}\mathop{\mathbb{E}}_{\mathbf{x}\sim q_{\theta}(\mathbf{x})}f(\mathbf{x}; \theta) = &\nabla^2_{\theta}\mathop{\mathbb{E}}_{\epsilon \sim \pi(\epsilon)}f(h_{\theta}(\epsilon); \theta)\\
      = & \mathop{\mathbb{E}}_{\epsilon \sim \pi(\epsilon)} \nabla^2_{\theta}f(h_{\theta}(\epsilon); \theta).
    \end{split}
\end{equation}

\subsection{First- and second-order REINFORCE estimation}
The REINFORCE estimation utilizes the log-derivative trick which is
\begin{equation}
  \nabla_{\theta}q_{\theta}(\mathbf{x}) = q_{\theta}(\mathbf{x}) \nabla_{\theta}\log q_{\theta}(\mathbf{x}).
\end{equation}
The first-order derivative is
\begin{equation}
  \begin{split}
    & \nabla_{\theta}\mathop{\mathbb{E}}_{\mathbf{x}\sim q_{\theta}(\mathbf{x})}f(\mathbf{x}; \theta) =  \nabla_{\theta} \sum_{\{\mathbf{x}\}} q_{\theta}(\mathbf{x}) f(\mathbf{x}; \theta)\\
    = & \sum_{\{\mathbf{x}\}} \left[ f(\mathbf{x}; \theta) q_{\theta}(\mathbf{x}) \nabla_{\theta}\log q_{\theta}(\mathbf{x}) + q_{\theta}(\mathbf{x})\nabla_{\theta}f(\mathbf{x}; \theta)\right]\\
    = & \mathop{\mathbb{E}}_{\mathbf{x}\sim q_{\theta}(\mathbf{x})} \left[f(\mathbf{x}; \theta)\nabla_{\theta}\log q_{\theta}(\mathbf{x}) +  \nabla_{\theta}f(\mathbf{x}; \theta)\right].
  \end{split}
  \label{eq:reinforce1}
\end{equation}
Note that
\begin{equation}
  \mathop{\mathbb{E}}_{\mathbf{x}\sim q_{\theta}(\mathbf{x})}\nabla_{\theta}\log q_{\theta}(\mathbf{x}) = \nabla_{\theta}\sum_{\{\mathbf{x}\}}q_{\theta}(\mathbf{x})=  \nabla_{\theta} 1 =0.
\end{equation}
The variance of estimates of Eq.~(\ref{eq:reinforce1}) can thus be reduced via:~\cite{reparamAndreinforce}
\begin{equation}
  \begin{aligned}
 & \nabla_{\theta}\mathop{\mathbb{E}}_{\mathbf{x}\sim q_{\theta}(\mathbf{x})}f(\mathbf{x}; \theta)\\
  =  &\mathop{\mathbb{E}}_{\mathbf{x}\sim q_{\theta}(\mathbf{x})}\left[(f(\mathbf{x}; \theta) - b )\nabla_{\theta}\log q_{\theta}(\mathbf{x})+  \nabla_{\theta}f(\mathbf{x}; \theta)\right],
  \end{aligned}
\end{equation}
where $b$ is the baseline. The baseline value $b$ has many different formulas \cite{reparamAndreinforce} and here we used the simplest one:
\begin{equation}
  b = \mathop{\mathbb{E}}_{\mathbf{x}\sim q_{\theta}(\mathbf{x})}f(\mathbf{x}; \theta).
\end{equation}
The second-order derivative is
\begin{equation}
  \begin{aligned}
    &\nabla^2_{\theta}\mathop{\mathbb{E}}_{\mathbf{x}\sim q_{\theta}(\mathbf{x})}f(\mathbf{x}; \theta) \\
    =& \nabla_{\theta} \sum_{\{\mathbf{x}\}}  f(\mathbf{x}; \theta) q_{\theta}(\mathbf{x}) \nabla_{\theta}\log q_{\theta}(\mathbf{x}) + q_{\theta}(\mathbf{x})\nabla_{\theta}f(\mathbf{x}; \theta) \\
    =&\sum_{\{\mathbf{x}\}} f(\mathbf{x}; \theta)q_{\theta}(\mathbf{x})\nabla^2_{\theta}\log q_{\theta}(\mathbf{x}) + f(\mathbf{x}; \theta)q_{\theta}(\mathbf{x})(\nabla_{\theta}\log q_{\theta}(\mathbf{x}))^2 \\
     & + \sum_{\{\mathbf{x}\}} q_{\theta}(\mathbf{x})\nabla_{\theta}^2f(\mathbf{x}; \theta) +2 \nabla_{\theta}f(\mathbf{x}; \theta)q_{\theta}(\mathbf{x}) \nabla_{\theta}\log q_{\theta}(\mathbf{x}) \\
    = &\mathop{\mathbb{E}}_{\mathbf{x}\sim q_{\theta}(\mathbf{x})} \Big[ f(\mathbf{x}; \theta) \left(\nabla^2_{\theta}\log q_{\theta}(\mathbf{x}) +\left(\nabla_{\theta}\log q_{\theta}(\mathbf{x})\right)^2\right) \Big] \\
    & + \mathop{\mathbb{E}}_{\mathbf{x}\sim q_{\theta}(\mathbf{x})} \Big[ \nabla_{\theta}^2f(\mathbf{x}; \theta) + 2\nabla_{\theta}f(\mathbf{x}; \theta)\nabla_{\theta}\log q_{\theta}(\mathbf{x})  \Big].
  \end{aligned}
\end{equation}
Note that 
\begin{equation}
  \begin{aligned}
  &\mathop{\mathbb{E}}_{\mathbf{x}\sim q_{\theta}(\mathbf{x})} \left[ \nabla^2_{\theta}\log q_{\theta}(\mathbf{x}) +(\nabla_{\theta}\log q_{\theta}(\mathbf{x}))^2 \right] \\
  &= \nabla_{\theta}^{2}\sum_{\{\mathbf{x}\}} q_{\theta}(\mathbf{x}) = 0.
  \end{aligned}
\end{equation}
The variance reduction method can also be employed as 
\begin{equation}
  \begin{aligned}
 &\nabla^2_{\theta}\mathop{\mathbb{E}}_{\mathbf{x}\sim q_{\theta}(\mathbf{x})} f(\mathbf{x}; \theta)\\
  = &\mathop{\mathbb{E}}_{\mathbf{x}\sim q_{\theta}(\mathbf{x})} \Big[ ( f(\mathbf{x}; \theta) -b_1 )\left(\nabla^2_{\theta}\log q_{\theta}(\mathbf{x}) +\left(\nabla_{\theta}\log q_{\theta}(\mathbf{x})\right)^2\right) \Big]\\
  & +  \mathop{\mathbb{E}}_{\mathbf{x}\sim q_{\theta}(\mathbf{x})} \Big[ \nabla_{\theta}^2f(\mathbf{x}; \theta) +2 \left(\nabla_{\theta}f(\mathbf{x}; \theta) - b_2\right)\nabla_{\theta}\log q_{\theta}(\mathbf{x}) \Big] ,
  \end{aligned}
\end{equation}
where
\begin{equation}
  \begin{split}
  b_1 = &\mathop{\mathbb{E}}_{\mathbf{x}\sim q_{\theta}(\mathbf{x})} f(\mathbf{x}; \theta), \\
  b_2 = &\mathop{\mathbb{E}}_{\mathbf{x}\sim q_{\theta}(\mathbf{x})}  \nabla_{\theta}f(\mathbf{x}; \theta).
  \end{split}
\end{equation}

\subsection{Reparameterization of acceptance-rejection sampling}
The acceptance-rejection sampling is a common way to draw samples from certain relatively complex distributions, such as the von Mises distribution.
The reparameterization of acceptance-rejection sampling is given in Ref.~\cite{acptReparam}.
For a better illustration, we first give the general framework of acceptance-rejection sampling.

In acceptance-rejection sampling, to sample distribution $q_{\theta}(z)$, we first draw a random variable $\epsilon$ from the distribution $s(\epsilon)$.
It is then transformed using a parameterized transformation $h_{\theta}$,
similar to the NF, we should consider the Jacobian of this transformation as the change of the probability distribution. Then we note its changed possibility as $r_{\theta}({z})$, where ${z} = h_{\theta}(\epsilon)$.
The sample is accepted with the probability $\min\{1, \frac{q_{\theta}({z})}{M_{\theta}r_{\theta}({z})}\}$ or resampled till being accepted, as in Alg.~\ref{alg:rejectionSampling}.

  \begin{algorithm}[H]
    \caption{acceptance-rejection sampling}
    \label{alg:rejectionSampling}
    \SetKwInOut{Input}{Input}
    \SetKwInOut{Output}{Output}
    \Input{$s$, $h_{\theta}$, $M_{\theta}$, and $q_{\theta}$}
    \Output{${z}=h_{\theta}(\epsilon)$}
    $i \leftarrow 0$\;
    \Repeat{$u < \frac{q_{\theta}({z})}{M_{\theta}r_{\theta}({z})}$}{
     $i \leftarrow i+1$\;
     $ \epsilon \sim s(\epsilon), {z}=h_{\theta}(\epsilon)$\;
     $u \sim {U}(0,1)$\;
    }
    \Return ${z}$
\end{algorithm}  

The probability distribution of the accepted $\epsilon$ is
\begin{equation}
  \begin{aligned}
  \pi_{\theta}(\epsilon)
  = &\int M_{\theta} s(\epsilon) \mathds{1}\left(0 < u < \frac{q_{\theta}(h_{\theta}(\epsilon))}{M_{\theta}r(h_{\theta}(\epsilon); \theta)}\right) d u
  \\= & s(\epsilon)\frac{q_{\theta}(h_{\theta}(\epsilon))}{r(h_{\theta}(\epsilon); \theta)}.
  \end{aligned}
\end{equation}
Adapting the reparameterization, the first-order derivative can be reformulated as
\begin{equation}
  \nabla_{\theta} \mathop{\mathbb{E}}_{{z}\sim q_{\theta}({z})}f({z} ;{\theta}) = \nabla_{\theta} \mathop{\mathbb{E}}_{\epsilon \sim \pi_{\theta}(\epsilon)}f(h_{\theta}(\epsilon); \theta).
\end{equation}
Then, using the log-derivative trick of REINFORCE, this becomes
\begin{equation}
   \mathop{\mathbb{E}}_{\epsilon \sim s(\epsilon)}\left[ \frac{q_{\theta}(h_{\theta}(\epsilon))}{r(h_{\theta}(\epsilon); \theta)} \nabla_{\theta} f(h_{\theta}(\epsilon); \theta)\right] + \mathop{\mathbb{E}}_{\epsilon \sim s(\epsilon)}\left[ \frac{q_{\theta}(h_{\theta}(\epsilon))}{r(h_{\theta}(\epsilon); \theta)} f(h_{\theta}(\epsilon); \theta) \nabla_{\theta} \log \frac{q_{\theta}(h_{\theta}(\epsilon))}{r(h_{\theta}(\epsilon); \theta)} \right].
\end{equation}

We generalize the scheme in Ref.~\cite{acptReparam}, and give the second-order derivative
\begin{equation}
    \nabla_{\theta}^2 \mathop{\mathbb{E}}_{{z}\sim q_{\theta}({z})}f({z};{\theta}) = \nabla_{\theta}^2 \mathop{\mathbb{E}}_{\epsilon \sim \pi_{\theta}(\epsilon)}f(h_{\theta}(\epsilon); \theta).
  \end{equation}
  This can be reformulated as
  \begin{equation}
    \begin{aligned}
      &\nabla_{\theta} \int s(\epsilon) \frac{q_{\theta}(h_{\theta}(\epsilon))}{r(h_{\theta}(\epsilon); \theta)} \nabla_{\theta} f(h_{\theta}(\epsilon); \theta) d \epsilon
    \\& + \nabla_{\theta} \int s(\epsilon)f(h_{\theta}(\epsilon); \theta) \nabla_{\theta} \frac{q_{\theta}(h_{\theta}(\epsilon))}{r(h_{\theta}(\epsilon);\theta)} d \epsilon
    \\ = &2\mathop{\mathbb{E}}_{\epsilon \sim s(\epsilon)}\left[ \frac{q_{\theta}(h_{\theta}(\epsilon))}{r(h_{\theta}(\epsilon);\theta)} \nabla_{\theta}f(h_{\theta}(\epsilon); \theta) \nabla_{\theta} \log \frac{q_{\theta}(h_{\theta}(\epsilon))}{r(h_{\theta}(\epsilon);\theta)} \right] \\
    &+ \mathop{\mathbb{E}}_{\epsilon \sim s(\epsilon)}\left[ \frac{q_{\theta}(h_{\theta}(\epsilon))}{r(h_{\theta}(\epsilon); \theta)} \nabla_{\theta}^2 f(h_{\theta}(\epsilon); \theta)\right]\\
    &+ \mathop{\mathbb{E}}_{\epsilon \sim s(\epsilon)}\left[ f(h_{\theta}(\epsilon); \theta) \nabla_{\theta}^2  \frac{q_{\theta}(h_{\theta}(\epsilon))}{r(h_{\theta}(\epsilon); \theta)} \right].
    \end{aligned}
\end{equation}
Note that
\begin{equation}
\begin{aligned}
  &\mathop{\mathbb{E}}_{\epsilon \sim s(\epsilon)} \nabla_{\theta} \frac{q_{\theta}(h_{\theta}(\epsilon))}{r(h_{\theta}(\epsilon); \theta)} = \int \nabla_{\theta} \pi_{\theta}(\epsilon) d \epsilon= 0,\\
  &\mathop{\mathbb{E}}_{\epsilon \sim s(\epsilon)} \nabla_{\theta}^2 \frac{q_{\theta}(h_{\theta}(\epsilon))}{r(h_{\theta}(\epsilon); \theta)} = \int \nabla_{\theta}^2 \pi_{\theta}(\epsilon) d \epsilon= 0,
\end{aligned}
\end{equation}
so the variance reduction method \cite{reparamAndreinforce} can also be applied here. 

For the von Mises distribution $\text{VM}(\mu, \kappa)$, the $s$ distribution is
\begin{equation}
 \epsilon_1, \epsilon_2 \sim \text{U}(0, 1).
\end{equation}
The transformation $h_{\theta}$ is
\begin{equation}
  h_{\theta}(\epsilon_1, \epsilon_2) = \text{sign} (\epsilon_2 - 0.5) \cos^{-1}\left(\frac{1+cz}{c+z}\right) + \mu,
\end{equation}
where
\begin{equation}
  \begin{aligned}
    & b = 1 + \sqrt{1 + 4\kappa^2},\\
    &\rho = (b - \sqrt{2b}) / 2\kappa,\\
    &c = (1 + \rho^2) / (2 \rho), \\
    &z = \cos (\pi \epsilon_1). \\
  \end{aligned}
\end{equation}
Similar to the NF model, the $r_{\theta}$ distribution can be derived using the $s$ distribution and the Jacobian of $h_{\theta}$.

\section{Thermodynamic Quantities from the Differentiation of the Free Energy}

\subsection{Theoretical analysis of the derivative estimation method compared to the statistical averaging with direct sampling}

After training a VaTD model, in addition to the derivative estimation method used in the experiment (as shown in Eq. (\ref{eq:diffmeanEandCV})), an alternative approach to estimating mean energy and heat capacity is to perform statistical averaging with direct sampling, \ie,
\begin{equation}
  \langle E \rangle = \mathop{\mathbb{E}}_{\mathbf{x}\sim q_{\theta}} E(\mathbf{x}), ~ C_v = \beta^2(\mathop{\mathbb{E}}_{\mathbf{x}\sim q_{\theta}} E^2(\bm{x}) - (\mathop{\mathbb{E}}_{\mathbf{x}\sim q_{\theta}} E(\bm{x}))^2).
\end{equation}
For a better illustration, we first establish a connection between the two methods.

Using Eq.~(\ref{eq:diffmeanEandCV}), given the estimated $-\log\bar{\mathcal{Z}}(\beta)$, the mean energy is
\begin{equation}
  \begin{aligned}
    \langle E \rangle = & -\frac{\partial \log \bar{\mathcal{Z}}(\beta)}{\partial \beta}\\
    = & \frac{\partial }{\partial \beta}\sum_{\mathbf{x}}q_{\theta}(\mathbf{x}, \beta)\left[ \log q_{\theta}(\mathbf{x}, \beta) + \beta E(\mathbf{x})
    \right]\\
    = & \sum_{\mathbf{x}} q_{\theta}(\mathbf{x}, \beta) \left[\log q_{\theta}(\mathbf{x}, \beta) + \beta E(\mathbf{x}) \right]\frac{\partial \log q_{\theta}(\mathbf{x}, \beta)}{\partial \beta} \\
    & +\sum_{\mathbf{x}}q_{\theta}(\mathbf{x}, \beta)E(\mathbf{x}).
    \end{aligned}
    \label{eq:meanE}
\end{equation}
It can be seen that if the learned distribution is the exact Boltzmann distribution, \ie,
\begin{equation}
  q_{\theta}(\mathbf{x}, \beta) \vcentcolon= \frac{\exp(-\beta E(\mathbf{x}))}{\mathcal{Z}},
  \label{eq:BoltzmannDist}
\end{equation}
Eq.~(\ref{eq:meanE}) becomes
\begin{align}
  \langle E \rangle =\mathop{\mathbb{E}}_{\mathbf{x}\sim q_{\theta}(\mathbf{x}, \beta)} E(\mathbf{x}).
\end{align}

A similar analysis can be done for the heat capacity with the same conclusion; that is,
when the training is perfectly done, the derivative estimation method gives the same results as those obtained from statistical averaging.

For imperfect training, the two methods give different results. The derivative estimation, which is closely related to the free energy (the loss function), typically has better accuracy compared to statistical averaging.
Additionally, one can view the VaTD model as a generalization of the variational mean-field method.
In the mean-field theory, when the variational ansatz $q_{\theta}$ is not flexible enough, the $q_{\theta}$ does not converge to the exact distribution, but still can have a relatively accurate estimate of the free energy and partition function.

On the other hand, even with perfect or near-perfect training, it is better to choose the derivative estimation method.
This is because the derivative estimation method has a much faster convergence speed;
from Eq.~(\ref{eq:KLD}), one can see that a perfect model will give uniformly the same estimate of the free energy for any $\bm{x}$.
In this sense, a batch of only one sample will converge.

We further give numerical tests in the following sections to demonstrate these two advantages of the derivative estimation method.

\subsection{Theoretical analysis of the convergence time of derivative estimation}
\change{

 In VaTD models, when estimating the mean energy or magnetization,
  each sample is i.i.d. sampled, so the estimates are also i.i.d., which means the convergence error $\sigma_N$ follows the variance of the mean of $N$ independent values each with variance $\sigma^2$, \ie,
  \begin{equation}
    \sigma_{N} = \sqrt{\frac{\sigma^2}{N}}
    \label{eq:convT}
  \end{equation}
  In MCMC, where samples are correlated, the estimates from each sample are also correlated.
  Then, as a simple analysis, one can use a $\Delta t$ as the time step interval to take samples, which should be much larger than the autocorrelation time $\tau$ to make sure that each sample closes to an i.i.d. sample.
  This means the effective sample batch size is $N_{eff} = N / \Delta t$, \ie, the autocorrelation time introduces a scaling factor that slows down the convergence.
  At PT, the correlation effect will dominate, where large autocorrelation times are observed.
  So, i.i.d. samplers, such as the proposed VaTD model, converge faster than MCMC, especially at PT.

  In the estimation of heat capacity, for MCMC, convergence is harder.
  This is because, the heat capacity is expressed as the deviation of energy in a batch of sampled states, which is a macroscopic statistical quantity and inevitably correlated to all samples in the batch.
  This means 
  heat capacity will converge slower than, for example, mean energy or magnetization.
  A simple convergence analysis like Eq.~(\ref{eq:convT}) will not apply. 
  More advanced methods, like the bootstrap method~\cite{bootstrap}, is required.
  For tractable density models with only a single fixed temperature, like \cite{prl} and \cite{wuDianPRL}, this is also the case, as they're also sample-based.
  In contrast, for VaTD models, by making the temperature differentiable, one can estimate the heat capacity via Eq.~(\ref{eq:diffmeanEandCV}).
  This is a fundamentally different approach.
  As in Eq.~(\ref{eq:diffmeanEandCV}), each i.i.d. sample gives an i.i.d. estimate of the free energy and thus an i.i.d. estimate of the heat capacity.
  So, the final estimate is still a mean over a batch of i.i.d. estimates, where Eq.~(\ref{eq:convT}) still applies and a faster convergence can be achieved.

  In the following section, we provide a convergence time comparison for experimental demonstrations.

}

\subsection{Numerical comparisons of convergence times of mean energy and heat capacity}
 \begin{table}[h!]
  \centering
\begin{tabular}{ |c||c|c|c|c|c|c|c|c|}
 \hline
  &\multicolumn{8}{|c|}{Temperature ($\beta$)} \\
 Methods & 0.65 &0.70 &0.90 & 0.91 & 0.95 & 1.15 & 1.20 & 1.25\\
 \hline\hline
  HMC ($\langle E \rangle$) &497.46 & 776.07&  526.67 & 518.43 & 435.43 & 452.98 & 469.50 & 568.46\\
  \hline
  VaTD ($\langle E \rangle$)&7.03  & 9.69& 38.94 &35.98 &23.61 & 8.24 & 6.70 & 6.39 \\
  \hline
  HMC ($C_v$)   & 3164.93    &3959.81&  21863.21 & 11719.66 & 19654.64 & 5575.14 & 5495.62 &4883.55\\
  \hline
  VaTD ($C_v$)&   35.98  &60.19    &256.01 & 243.72 &206.86 &146.40 & 143.21 & 142.83\\
 \hline
\end{tabular}
\caption{Comparison of convergence times of Hamiltonian Monte Carlo (HMC) and VaTD model in estimating heat capacity (\(C_v\)) and mean energy (\(\langle E \rangle\)) of the \(16 \times 16\) 2D XY model. 
The convergence threshold is \(10^{-3}\) for the mean energy (\(\langle E \rangle\)) and heat capacity (\(C_v\)) estimations. For the MCMC algorithm, we used a PyTorch implementation of HMC to leverage GPU parallelization and automatic differentiation for force calculations. For the VaTD model, we employed the same model presented in the manuscript.
All the tests are conducted using a same NVIDIA GeForce RTX 4090 GPU card.
The time is measured in seconds (s).}
\label{tab:time}
\end{table}

\begin{figure*}[!tp]
  \centering
    \includegraphics[width=1\linewidth]{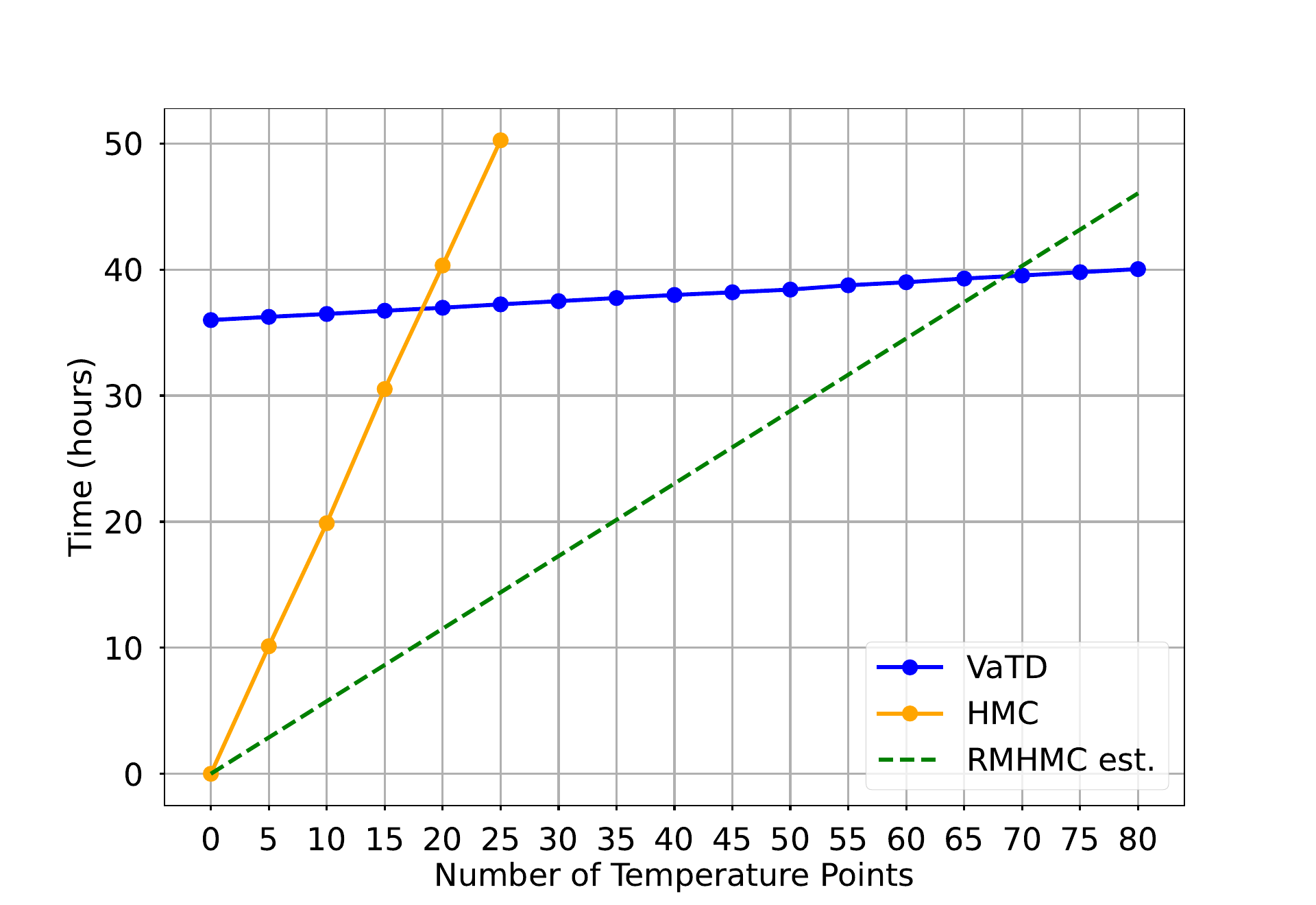}
  \caption{ 
  A comparison of total time consumption between VaTD and MCMC methods on a 2D $16\times16$ XY model with periodic boundary conditions. The temperature points are equally spaced within the range of $\beta \in [0.5, 1.5]$. The convergence accuracy is set at $10^{-3}$. We estimated the RMHMC time consumption curve using the speedup of $3.5$  relative to HMC, which is a upper bound of the reported best-case speedup of $3$.
}
\label{fig:timeCompare}
\end{figure*}
\begin{figure*}[bp]
  \centering
    \includegraphics[width=1\linewidth]{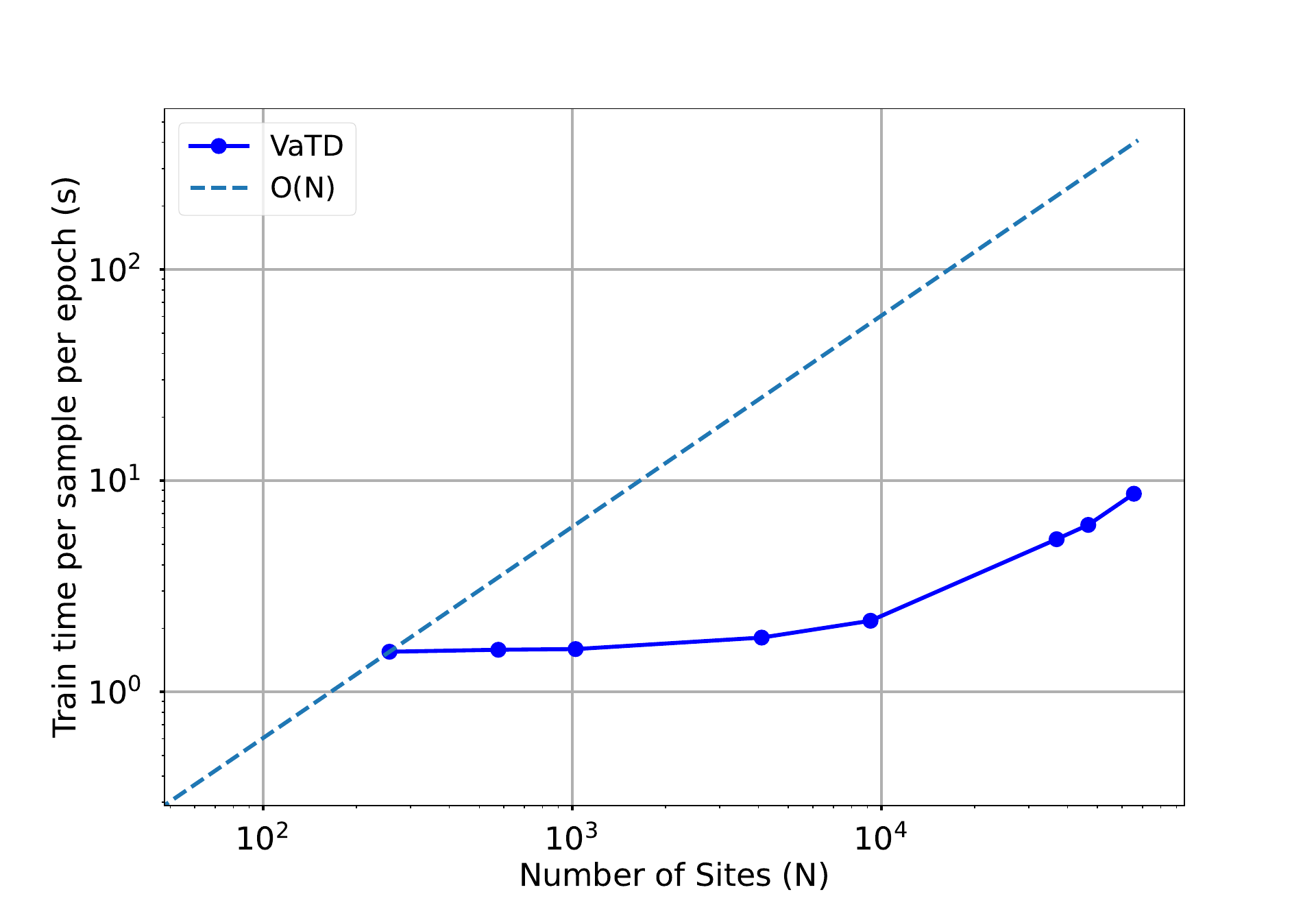}
  \caption{
  The mean training time per epoch per sample of 2D XY models as a function of the site number. The $O(N)$ line is estimated using the time data from the training of the $16\times16$ 2D XY model.
}
\label{fig:scalingXYTime}
\end{figure*}
\begin{figure*}[tb]
  \centering
    \includegraphics[width=1\linewidth]{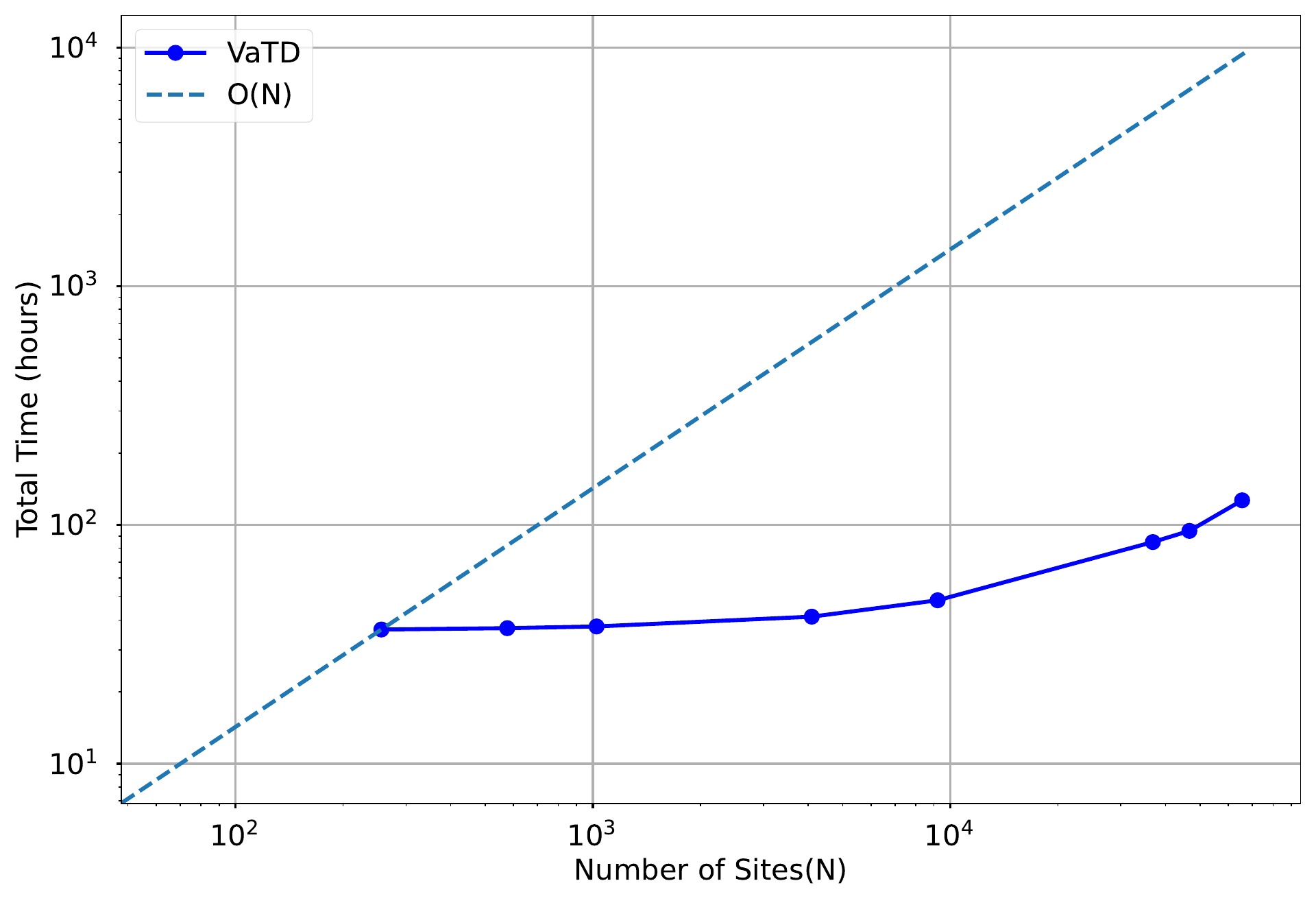}
  \caption{The total time of training and evaluating $10$ temperature points of 2D XY models as a function of the site number. The $O(N)$ line is estimated using the time data from the $16\times16$ 2D XY model. The convergence error is set to be $10^{-3}$.
}
\label{fig:totaltimeCompare}
\end{figure*}

\change{
In addition to the total time comparison in the main text,
  we also compared the convergence times of reaching an error of $10^{-3}$ for the VaTD model and Hamiltonian Monte Carlo (HMC).
  The results are shown in Tab.~\ref{tab:time}, and the time unit is second (s).
  
  For HMC, we used a batch size of $4000$, and a time step interval $\Delta t$ of $75$.
  The burn-in step of heat capacity was $1000$ while mean energy used a burn-in step of $20$.
  The integration algorithm in HMC is the leapfrog with Verlet time step.
  We also implemented an adaptive integration step size, which will keep the acceptance ratio above $80\%$.
  For error estimation in HMC, we used the bootstrap method \cite{bootstrap}.
  For the VaTD model, we used a batch size of $540$.
  And we used Eq.~(\ref{eq:diffmeanEandCV}) to estimate mean energy and heat capacity.
  For error estimation, as analysised above, we used Eq.~(\ref{eq:convT}).
  This convergence test is done on $8$ different temperatures, as listed in Tab.~\ref{tab:time}.

  These experimental results accurately reflects the theoretical analysis above.
  Firstly, in the convergence time comparison, VaTD model is generally $10 - 90$ times faster than HMC.
  As explained above, this is mostly due to the direct sampling of i.i.d. samples.
  Then, at the same temperature, a greater speedup from VaTD is observed in the convergence of heat capacity than mean energy.
  This is because
  VaTD can gain extra speedups by expressing heat capacity as a derivative of the free energy, as in Eq.~(\ref{eq:diffmeanEandCV}).
}

\changeT{
We further investigated the total time consumption of VaTD compared with MCMC methods in Fig.~\ref{fig:timeCompare}. 
In this figure, the proposed VaTD model included an initial overhead for training time, while HMC started from the origin.
The tested temperature points are equally spaced within the range ($\beta\in[0.5, 1.5]$), and we use a convergence accuracy of $10^{-3}$.
Fig.~\ref{fig:timeCompare} shows that VaTD scales more efficiently than HMC, implying a cross point is inevitable.
In this plot, the cross point happened at about $18$ temperature points.
As direct sampling gives i.i.d. samples without the effect of autocorrelation, this cross point will happen earlier at a lower convergence accuracy.

We also estimated the speed of the Riemann manifold HMC (RMHMC) in this comparison.
As reported in \cite{RMHMC}, the speedup of RMHMC over HMC ranges from $0.7$ to $3$, depending on the specific sampling datasets.
We thus estimated the time consumption curve of RMHMC using the speed ratio of $3.5$ relative to HMC, which is an upper bound of the best reported speedup $3$. In this best-case scenario, RMHMC would postpone the crossover by approximately $50$ additional temperature points.

In Fig.~\ref{fig:scalingXYTime}, we present a scaling test of training time as a function of different lattice sizes.
The results show that the training time scales efficiently, following a sub-$O(N)$ curve without significant slowdown.
Particularly, we observe a $6$-fold increase in average training time when the input dimension is increased by a factor of $100$. 
\changeTT{Similarly, in Fig.~\ref{fig:totaltimeCompare}, we also observed a sub-$O(N)$ curve when plotting the total time required for training and evaluating $10$ temperature points.}
This efficiency is attributed to the use of a CNN-based neural network architecture, which is inherently less sensitive to input dimensions.
}
\subsection{Estimation of the heat capacity using first-order derivative}

As in Eq.~(\ref{eq:diffmeanEandCV}), calculating the heat capacity requires the second-order derivative of the free energy. However, it is known that the second-order derivatives are not memory-efficient with backpropagation automatic differentiation. This can be easily resolved by the forward automatic differentiation. Nevertheless, backpropagation is more widely used, and as in the XY experiment we achieved very low STDs from the free energy estimates, we propose another scheme to estimate the heat capacity using only the first-order derivative.

Consider the first-order derivative of the log-probability,
\begin{equation}
  \begin{aligned}
  &\frac{\partial}{\partial \beta} \sum_{\{\mathbf{x}\}} q_{\theta}(\mathbf{x}, \beta) \log q_{\theta}(\mathbf{x}, \beta)\\
  =& \sum_{\{\mathbf{x}\}} \frac{\partial q_{\beta}(\mathbf{x}, \beta)}{\partial \beta} \log q_{\theta}(\mathbf{x}, \beta)  + \frac{\partial q_{\theta}(\mathbf{x}, \beta)}{\partial \beta} \\
  =& \sum_{\{\mathbf{x}\}} \frac{\partial q_{\theta}(\mathbf{x}, \beta)}{\partial \beta} \log q_{\theta}(\mathbf{x}, \beta),
  \end{aligned}
  \label{eq:Cv1st}
\end{equation}
where $q_{\theta}(\mathbf{x}, \beta)$ is the distribution of the VaTD model.
When the loss is low and STD is small, we assume the distribution of VaTD model has the form of the Boltzmann distribution, \ie,
\begin{equation}
  q_{\theta}(\mathbf{x}, \beta) \vcentcolon= \frac{\exp(-\beta E_{\theta}(\mathbf{x}))}{\bar{\mathcal{Z}}},
\end{equation}
where $E_{\theta}$ is the energy function learned by the VaTD model, and because of a small STD, $\bar{\mathcal{Z}}$ is a stable estimate of the partition function which doesn't depend on $\bm{x}$.
Then, Eq.~(\ref{eq:Cv1st}) can be reformulated into
\begin{equation}
  \begin{aligned}
  &\sum_{\{\mathbf{x}\}} q_{\theta}(\mathbf{x}, \beta) \left( -E_{\theta}(\mathbf{x}) - \frac{\partial \log \bar{\mathcal{Z}}}{\partial \beta}\right) (-\beta E_{\theta}(\mathbf{x}, \beta) - \log \bar{\mathcal{Z}}) \\
  = & \sum_{\{\mathbf{x}\}} q_{\theta}(\mathbf{x}, \beta) ( E_{\theta}(\mathbf{x}) - \langle E_{\theta} \rangle) (\beta E_{\theta}(\mathbf{x}, \beta) + \log \bar{\mathcal{Z}}) \\
  =& \sum_{\{\mathbf{x}\}} \left[q_{\theta}(\mathbf{x}, \beta) ( E_{\theta}(\mathbf{x}) - \langle E_{\theta} \rangle) \log \bar{\mathcal{Z}} \right] + \beta (\langle E^2_{\theta} \rangle - \langle E_{\theta} \rangle^2).
   \end{aligned}
   \label{eq:1stCvEst}
\end{equation}
As we consider $\log\bar{\mathcal{Z}}$ is a constant for any $\bm{x}$,
the former term in Eq.~(\ref{eq:1stCvEst}) disappears, and the latter is related to the heat capacity of the learned energy function.
So, the heat capacity estimate can be written as
\begin{equation}
  C_v = \beta \frac{\partial}{\partial \beta} \left[\mathop{\mathbb{E}}_{\mathbf{x}\sim q_{\theta}(\cdot, \beta)}\log q_{\theta}(\mathbf{x}, \beta)\right] \approx \beta^2 (\langle E^2_{\theta} \rangle -  \langle E_{\theta} \rangle^2),
\end{equation}
which can be estimated using only the first-order derivative.

\section{Supplementary for the 2D Ising Model Experiment}
\subsection{Architecture and hyperparameters}
The architecture we used is a standard ResNet \cite{resnet} structure with the CNN layer replaced by the masked CNN layer of Ref.~\cite{pixelRCNN}.
For training, we used the Adam optimizer \cite{adam}, whose learning rate was reduced periodically.
Because the REINFORCE estimation tends to have high variance, we clipped the gradient values that are greater than a maximum value.
All structural and training parameters can be found in Tab.~\ref{tab:paramIsing}.

\change{We used a collection of 8 temperature points at $\beta = 0.5, 0.8, 0.9, 1, 1.1, 1.2, 1.3, 1.8$ as the evaluation loss. We first trained the model till this evaluation loss reached equilibrium. From those saved models at equilibrium, we then selected the optimal model with the smallest maximum difference with the exact results over the temperature range of $[0.1, 1.0]$
}

\change{For the plots in the main text, we used Eq.~(\ref{eq:convT}) to control the convergence error to a level of $5\times 10^{-3}$.}

\begin{table}[htb]
  \centering
  \caption{Hyperparameters of the VaTD PixelCNN model used in the 2D Ising model experiment.}
  \label{tab:paramIsing}
  \begin{tabular}{c|c}
    \hline
    Structural parameter  &  \\
    \hline
    ResNet kernel size       &13   \\
    ResNet channel           &64  \\
    ResNet block             &6  \\
    Fully-connected CNN layer&2  \\
    Activation function      &RELU \cite{relu} \\
    \hline
    Training parameter &     \\
    \hline
    Batch size      & 500 \\
    Learning rate   & $5\times10^{-4}$ \\
    Decay factor    & 0.92  \\
    Gradient clip   & 1.0    \\
    Betas           &(0.9, 0.999) \\
    Eps             &$10^{-8}$ \\
    \hline
  \end{tabular}
\end{table}

\subsection{Statistical averages with direct sampling}
In this section, we compared the numerical results from the free energy derivative estimation method with those from the statistical averaging with direct sampling.
As the model used to plot Fig.~\ref{fig:ising} of the main text has a very low loss value, one may think the condition of Eq.~(\ref{eq:BoltzmannDist}) is nearly satisfied,
so a clear difference between these two methods is not visible.
To better illustrate it numerically, we used a saved model near the end of the training which has a slightly higher loss value.

\begin{figure}[tb]
  \centering
  \includegraphics[width=0.5\linewidth]{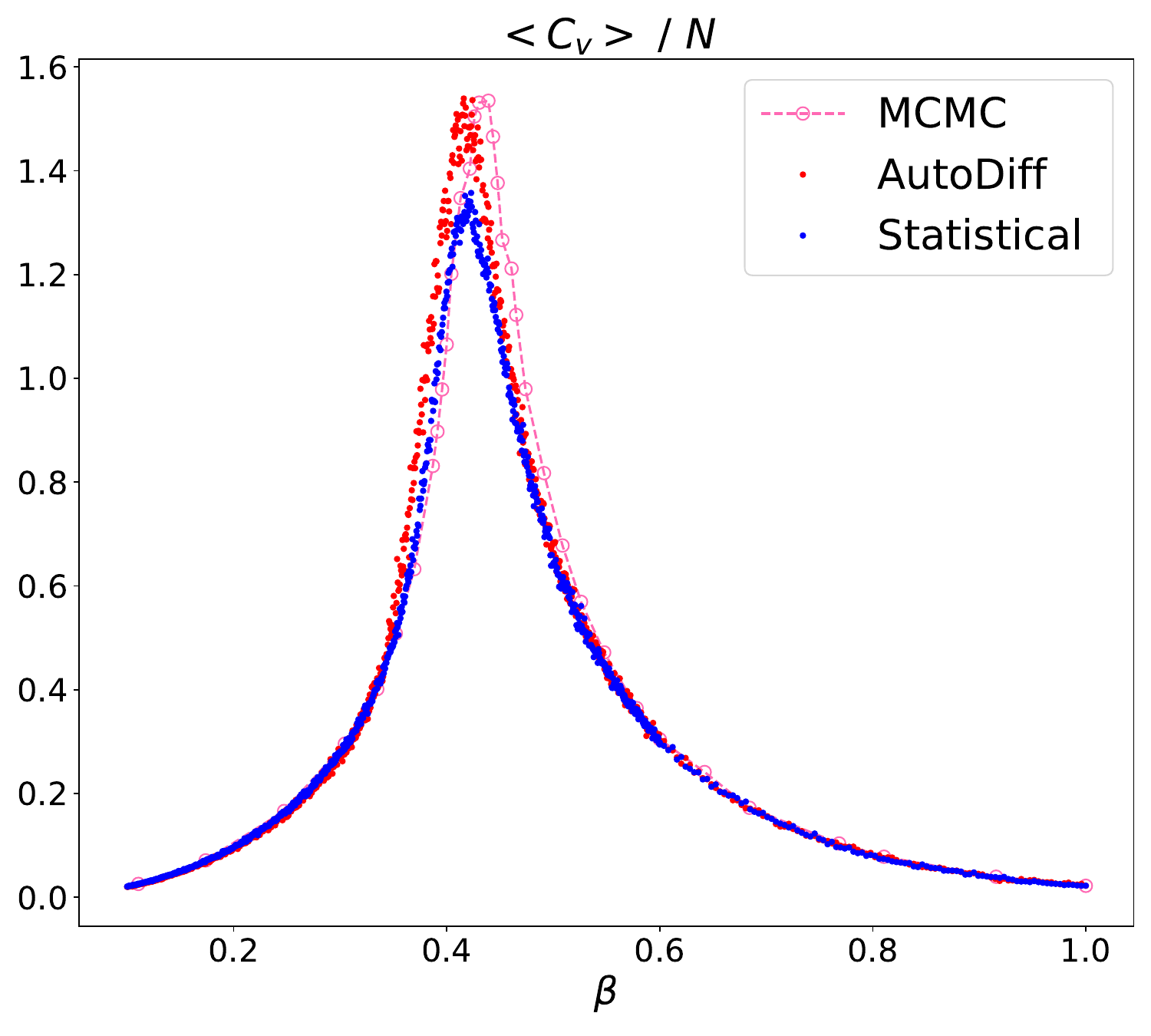}
  \caption{
    The estimated heat capacity of the 2D Ising model on a $16 \times 16$ square lattice with PBC.
    The estimation results using the differentiation of free energy (labeled as autoDiff) and statistical averaging with direct sampling (labeled as statistical) are compared with the MCMC results.
  }
  \label{fig:staIsing}
\end{figure}

In Fig.~\ref{fig:staIsing}, using this secondary model, we plotted the heat capacity values of the 2D Ising model using free energy derivative estimation and statistical averaging with direct sampling. Compared with the derivative estimation results, statistical averages show a larger deviation.
In this plot, for both the derivative estimation and statistical averaging, a batch of $7000$ samples was generated by the model at each temperature point.

\section{Supplementary for the 2D XY Model Experiment}

\subsection{Architecture and hyperparameters}

Our architecture consists of multiple coupling layers of the piece-wise cubic spline transformation \cite{cubic}.
The cubic spline transformation uses interpolation points to create a monotonically increasing function to parameterize an invertible transformation.
The coordinates and gradients of these interpolation points were computed using a standard ResNet \cite{resnet}.
This cubic spline flow also needs a way to bisect the variables.
As the 2D XY configuration is a 2D array, we used the checkerboard pattern to separate it into two parts with an equal amount of variables.
For training, we used the Adam optimizer \cite{adam}, and we decreased the learning rate periodically.
All structural and training parameters can be found in Tab.~\ref{tab:paramXY}.

\change{We used a collection of 8 temperature points at $\beta = 0.5, 0.75, 0.85, 0.9, 1., 1.1, 1.25, 2.4$ as the evaluation loss. We first trained the model till this evaluation loss reached equilibrium. From these saved models at equilibrium, we then selected the optimal model with the smallest maximum STD over the temperature range of $[0.5, 1.5]$.
}

\change{For the plots in the main text, we used Eq.~(\ref{eq:convT}) to control the convergence error to a level of $5\times 10^{-3}$.}

\begin{table}[htb]
  \centering
  \caption{Hyperparameters of the VaTD NF model used in the 2D XY model experiment.}
  \label{tab:paramXY}
  \begin{tabular}{c|c}
    \hline
    Structural parameter  &      \\
    \hline
    Cubic transformation layer & 5    \\
    Interpolation point number & 45   \\
    ResNet kernel size         &9     \\
    ResNet channel             &128   \\
    ResNet block               &6     \\
    Fully-connected CNN layer  &2     \\              
    Activation function        &ELU \cite{ELU} \\
    \hline
    Training parameter &  \\
    \hline
    Batch size         & 1024 \\
    Learning rate      & $7\times 10^{-4}$ \\
    Decay factor       & 0.7  \\
    Decay step         & 1000  \\
    Betas              &(0.9, 0.999) \\
    Eps                &$10^{-8}$ \\
    \hline
  \end{tabular}
\end{table}

\begin{figure}[tb]
  \centering
  \includegraphics[width=0.5\linewidth]{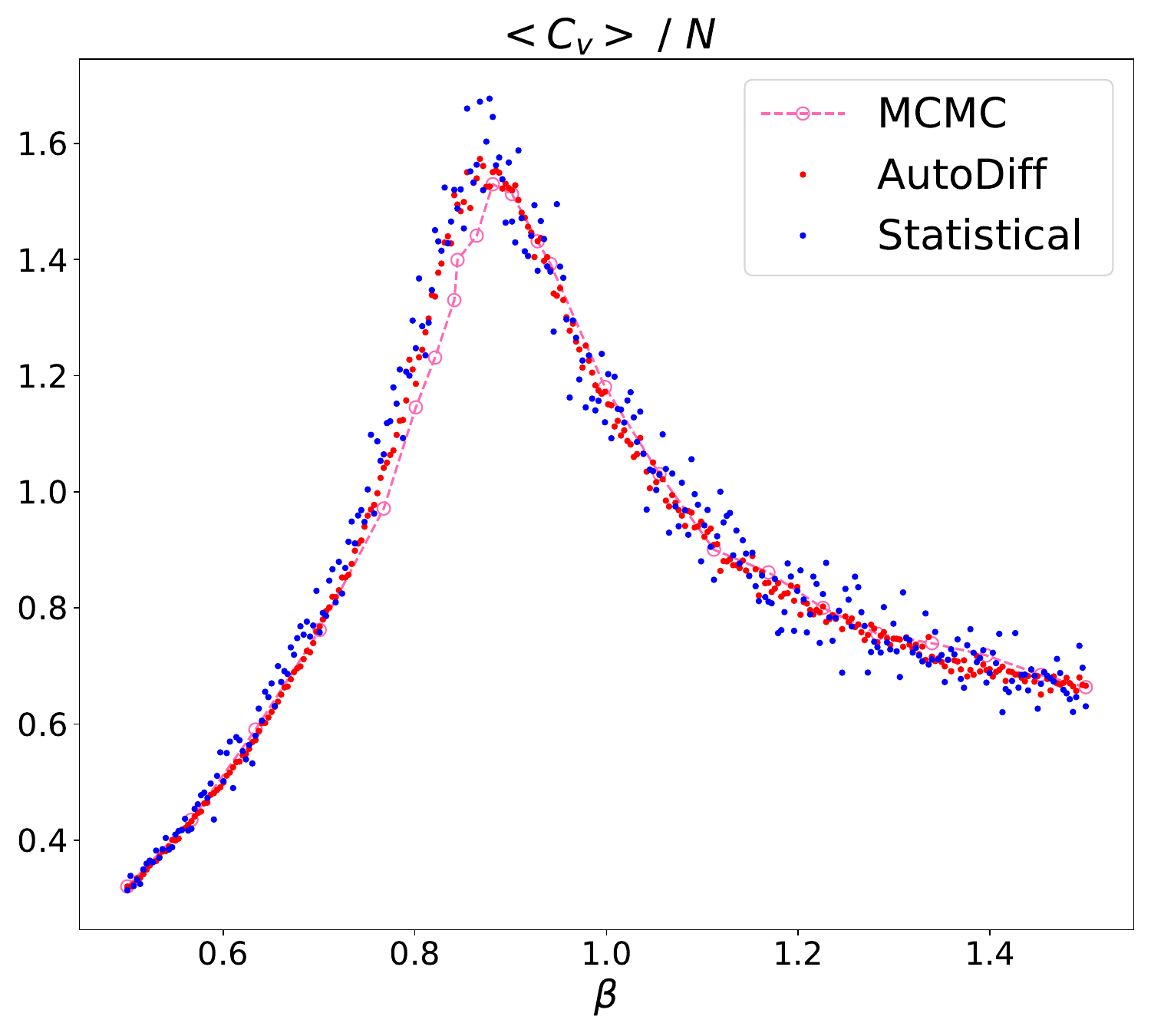}
  \caption{
    The estimated heat capacity of the 2D XY model on a $16 \times 16$ square lattice with PBC.
    The estimation results using the differentiation of free energy (labeled as autoDiff) and statistical averaging with direct sampling (labeled as statistical) are compared with the MCMC results.
  }
  \label{fig:statXY}
\end{figure}

\subsection{Statistical averages with direct sampling}

Using this XY case, we further demonstrated the difference between the free energy derivative estimation method and statistical averaging with direct sampling.
In Fig.~\ref{fig:statXY}, we gave the statistical averaging results of the heat capacity compared with derivative estimation results.
In this plot, for both the statistical averages and derivative estimation results, a batch of $1000$ samples was directly sampled by the model at each temperature point.
One can see that the statistical averages have a larger variance.

\subsection{Scaling test on larger lattice sizes}

\begin{figure*}[tb]
  \centering
    \includegraphics[width=1\linewidth]{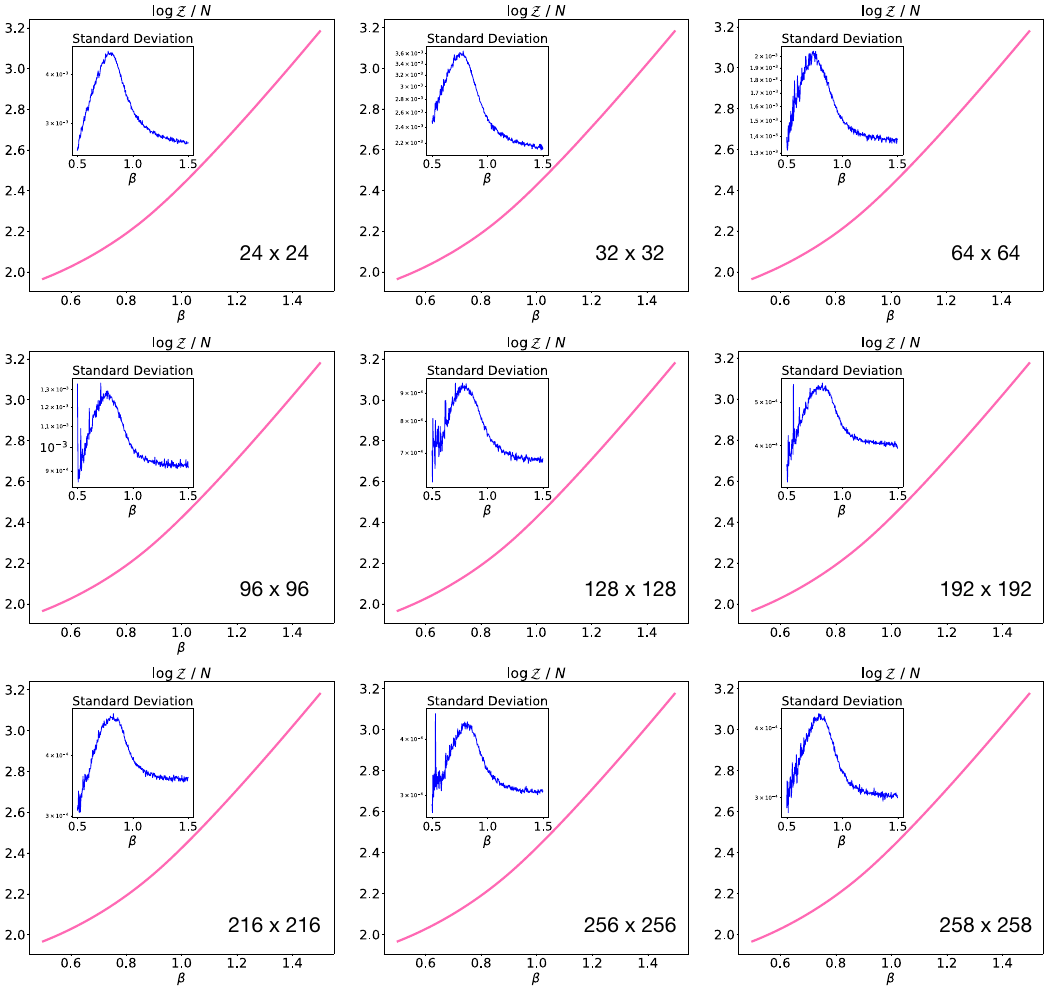}
 \caption{The estimated free energy and its standard deviation (inset) of the 2D XY model on a square lattice with periodic boundary conditions. The corresponding lattice sizes are labeled in the bottom-right corner of each subfigure. The temperature factor ($-T$) is removed from the free energy($-T \log \mathcal{Z}$) for better comparison.
}
\label{fig:scalingXYfreeE}
\end{figure*}
\begin{figure*}[tb]
  \centering
  \includegraphics[width=1\linewidth]{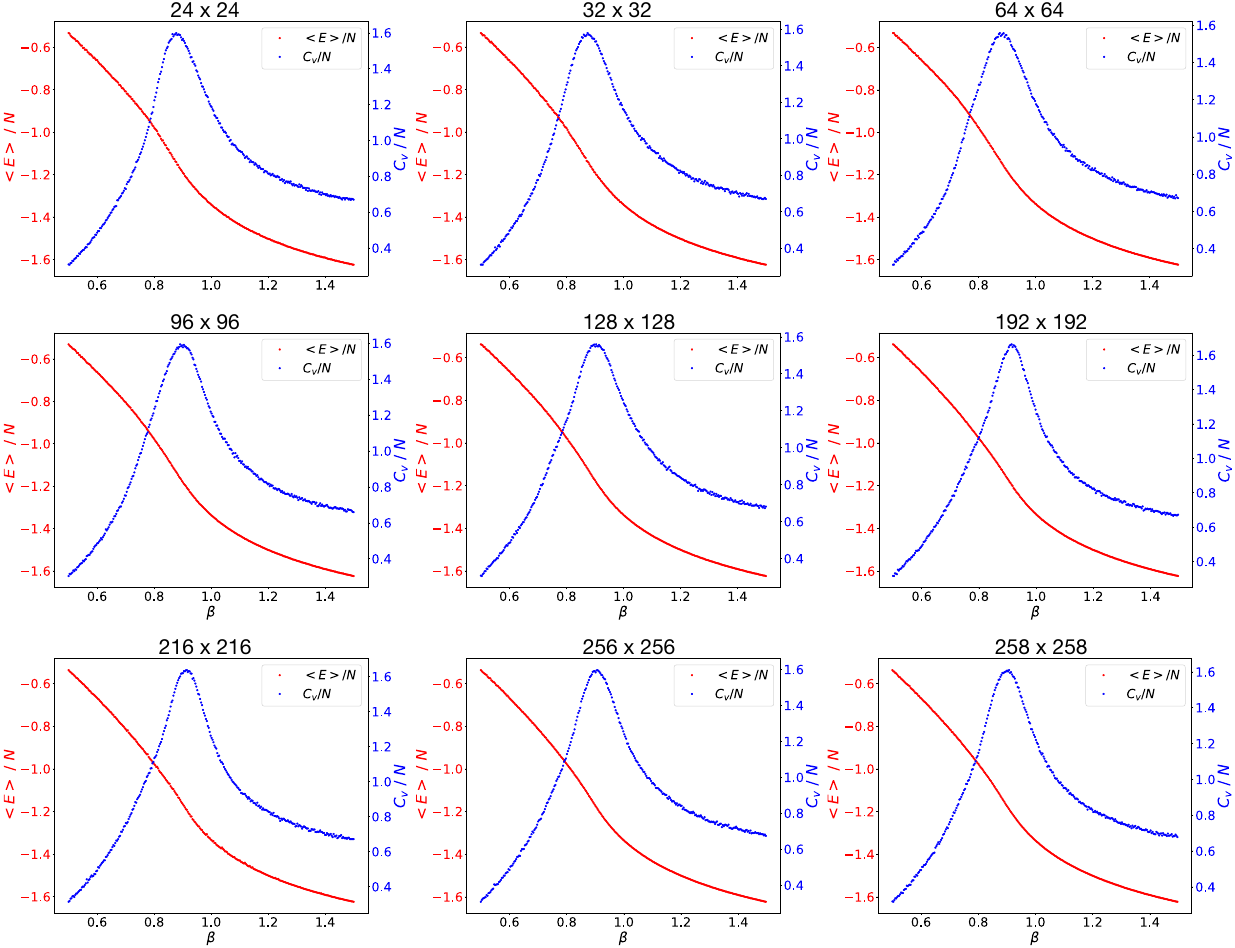}
  \caption{
The mean energy and heat capacity of the 2D XY model on a square lattice with periodic boundary conditions, estimated using the differentiation of the free energy.
The corresponding lattice sizes are labeled above each subfigure.
}   
  
\label{fig:scalingXY}
\end{figure*}

\change{
  To better understand the performance of the VaTD model used in the XY model experiment, we performed a scaling test on \changeT{$9$} other XY models with larger lattice sizes:
  \changeT{$24\times24$, $32\times32$, $64\times64$, $96\times96$, $128\times128$, $192\times192$, $216\times216$, $256\times256$, and $258\times258$.}
  In this scaling test, the same VaTD model used in the $16\times 16$ XY model experiment in the main text is used, \ie, with the same hyperparameters and the same structure as shown in Tab.~\ref{tab:paramXY}.
  \changeT{Due to the GPU memory limitation ($24$GB on RTX 4090), for lattice sizes larger than $32\times32$, we used checkpointing and reduced batch size.}

 \changeT{In Fig.~\ref{fig:scalingXYfreeE} and Fig.~\ref{fig:scalingXY}, we present plots of the free energy, mean energy, and specific heat capacity ($C_v$) for the 2D XY model at various lattice sizes.
These results demonstrate that the proposed VaTD training method can still identify phase transition phenomena in the 2D XY model up to a lattice size of $258\times258$.
We will need larger GPU memory or more GPU cards for a larger lattice size.
}

  We acknowledge that a fixed neural network model tends to exhibit diminished performance with larger input sizes. However,
  it is important to note that the proposed VaTD method serves as a general training framework that can integrate temperature dependence into \emph{any} tractable density generative model.
  For those targeting larger lattice sizes, we recommend using a more capable generative mode,
  such as the NeuralODE model~\cite{neuralode} or the continuous normalizing flow (CNF) models.
  Alternatively, generative models incorporating a renormalization group mechanism, as discussed in~\cite{prl, prr}.
  Both types of generative models can be effectively adopted
within the VaTD training framework.
}

\end{document}